%% file: main.tex
\documentclass[sigconf]{acmart}

\AtBeginDocument{%
  \providecommand\BibTeX{{%
    \normalfont B\kern-0.5em{\scshape i\kern-0.25em b}\kern-0.8em\TeX}}}
    
\setcopyright{acmcopyright}
\copyrightyear{2022}
\acmYear{2022}
\acmDOI{10.1145/1122445.1122456}

\acmConference[ICS '22]{ICS '22: ACM International Conference on Super Computing}{June 27--30, 2022}{}
\acmBooktitle{ICS '22: ACM International Conference on Super Computing,
  June 27--30, 2022, Held Virtually}
    
\usepackage{multirow}

\usepackage{graphicx}
\usepackage{textcomp}
\usepackage{xcolor}

\usepackage{multirow}
\usepackage{graphicx}

\copyrightyear{2022} 
\acmYear{2022} 
\setcopyright{acmcopyright}\acmConference[ICS '22]{2022 International Conference on Supercomputing}{June 28--30, 2022}{Virtual Event, USA}
\acmBooktitle{2022 International Conference on Supercomputing (ICS '22), June 28--30, 2022, Virtual Event, USA}
\acmPrice{15.00}
\acmDOI{10.1145/3524059.3532382}
\acmISBN{978-1-4503-9281-5/22/06}

\include{macros}

\begin{document}

\title{Parallel K-clique Counting on GPUs}

\input{authors}

\input{sec/0-abstract}



\settopmatter{printfolios=true}
\maketitle

\input{sec/1-intro}

\input{sec/2-background}

\input{sec/3-implementation}

\input{sec/4-evaluation}

\input{sec/5-related}

\input{sec/6-conculsion}

\input{sec/ack}

\balance
\bibliographystyle{ACM-Reference-Format}
\bibliography{ref}

\end{document}

%% file: macros.tex
\usepackage{balance}
\usepackage{algorithm}
\usepackage[noend]{algpseudocode}
\usepackage[normalem]{ulem}
\usepackage{pseudo}

\newcommand{\ignore}[1]{}

\iffalse
\newcommand{\todo}[1]{[{\color{red}\textbf{\sc TODO}: \textit{#1}}]}
\newcommand{\masri}[1]{[{\color{purple}\textbf{\sc MM}: \textit{#1}}]}
\newcommand{\izzat}[1]{[{\color{green}\textbf{\sc IE}: \textit{#1}}]}
\newcommand{\RN}[1]{[{\color{orange}\textbf{\sc RN}: \textit{#1}}]}
\newcommand{\JX}[1]{[{\color{red}\textbf{\sc JX}: \textit{#1}}]}
\newcommand{\change}[1]{{\color{blue}#1}}

\newcommand{\changerm}[1]{{\color{red}\sout{#1}}}
\else
\newcommand{\todo}[1]{}
\newcommand{\masri}[1]{}
\newcommand{\izzat}[1]{}
\newcommand{\RN}[1]{}
\newcommand{\JX}[1]{}
\newcommand{\change}[1]{#1}

\newcommand{\changerm}[1]{}
\fi

%% file: authors.tex
\author{Mohammad Almasri}
\email{almasri3@illinois.edu}
\affiliation{
    \institution{University of Illinois at Urbana-Champaign}
    \country{USA}
}

\author{Izzat El Hajj}
\email{izzat.elhajj@aub.edu.lb}
\affiliation{
    \institution{American University of Beirut}
    \country{Lebanon}
}

\author{Rakesh Nagi}
\email{nagi@illinois.edu}
\affiliation{
    \institution{University of Illinois at Urbana-Champaign}
    \country{USA}
}

\author{Jinjun Xiong}
\email{jinjun@buffalo.edu}
\affiliation{
    \institution{University at Buffalo}
    \country{USA}
}

\author{Wen-mei Hwu}
\email{w-hwu@illinois.edu}
\affiliation{
    \institution{University of Illinois at Urbana-Champaign, NVIDIA}
    \country{USA}
}

%% file: sec/0-abstract.tex
\begin{abstract}

Counting $k$-cliques in a graph is an important problem in graph analysis with many applications such as community detection and graph partitioning. 
Counting $k$-cliques is typically done by traversing search trees starting at each vertex in the graph.
Parallelizing $k$-clique counting has been well-studied on CPUs and many solutions exist.
However, there are no performant solutions for $k$-clique counting on GPUs.

Parallelizing $k$-clique counting on GPUs comes with numerous challenges such as the need for extracting fine-grain multi-level parallelism, sensitivity to load imbalance, and constrained physical memory capacity.
While there has been work on related problems such as finding maximal cliques and generalized sub-graph matching on GPUs, $k$-clique counting in particular has yet to be explored in depth.
In this paper, we present the first parallel GPU solution specialized for the $k$-clique counting problem.
Our solution supports both graph orientation and pivoting for eliminating redundant clique discovery.
It incorporates both vertex-centric and edge-centric parallelization schemes for distributing work across thread blocks, and further partitions work within each thread block to extract fine-grain multi-level parallelism while tolerating load imbalance.
It also includes optimizations such as binary encoding of induced sub-graphs and sub-warp partitioning to limit memory consumption and improve the utilization of execution resources.

Our evaluation shows that our best GPU implementation outperforms the best state-of-the-art parallel CPU implementation by a geometric mean of 12.39$\times$, 6.21$\times$, and 18.99$\times$ for $k=4$, $7$, and $10$, respectively.
We also perform a detailed evaluation of the trade-offs involved in the choice of parallelization scheme, and the incremental speedup of each optimization to provide an in-depth understanding of the optimization space.
The insights from our optimization flow can be useful for optimizing other clique finding and graph mining solutions on GPUs.
Our code will be open-sourced to enable further research on GPU parallelization of $k$-clique counting and other similar graph mining algorithms.

\end{abstract}

%% file: sec/1-intro.tex
\section{Introduction} \label{sec:intro}

Dense sub-graph counting and listing is an important problem in graph mining~\cite{dense2, dense3}.
A $k$-clique (or a $k$-vertex clique) in a graph is a complete sub-graph with exactly $k$ vertices and $k \times (k-1)$ edges, such that every vertex in the clique is connected to every other vertex.
Counting $k$-cliques is a useful algorithmic component of solutions to many problems such as community detection~\cite{cd1, cd2,  cd4, cd5}, graph partitioning and compression~\cite{gc1, gc2, gc3}, learning network embedding~\cite{ne1, ne2}, and recommendation systems~\cite{rs1, rs2}.

A common approach to $k$-clique counting is to traverse a search tree for each vertex and find $k$-cliques that contain that vertex.
This approach is commonly parallelized by processing different trees or subtrees in parallel.
One fundamental optimization is to eliminate the search tree branches that discover the same clique redundantly.
Two prominent approaches to this optimization are graph orientation~\cite{arb-count, kclist, chiba1985arboricity, orderingheuristic} and pivoting~\cite{powerpivoting}.

The graph orientation approach transforms the graph into a directed graph so that each $k$-clique, which is a symmetric structure, is only found from one of the vertices it contains.
Common orientation criteria include vertex degree~\cite{chiba1985arboricity, arb-count, degreeOrdering1}, graph coloring~\cite{orderingheuristic}, and degeneracy based on $k$-core decomposition~\cite{bkpivoting, kclist, arb-count} or relaxations of $k$-core~\cite{arb-count}.
The pivoting approach~\cite{powerpivoting} for $k$-clique counting is inspired by the Bron-Kerbosh maximal clique finding approach~\cite{bkpivoting}.
Rather than searching for all $k$-cliques, the pivoting approach finds the largest cliques, then calculates the number of $k$-cliques they contain.
To the best of our knowledge, the state-of-the-art parallel implementations for the graph orientation and pivoting approaches to $k$-clique counting are ARB-COUNT~\cite{arb-count} and Pivoter~\cite{powerpivoting}, respectively.
Both of these parallel implementations are designed for CPUs.

The massively parallel hardware resources in modern GPUs offer promising opportunities for accelerating $k$-clique counting for large graphs.
However, successful parallelization of $k$-clique counting on GPUs must overcome the additional challenges arising from the differences in hardware characteristics between GPUs and CPUs.
The first major challenge is that GPUs require more fine-grain parallelism to be extracted from the computation to utilize parallel hardware resources efficiently.
These parallel hardware resources are organized into a multi-level hierarchy which further complicates the parallelization process.
Moreover, the massively parallel nature of the hardware makes GPUs more sensitive to load imbalance.
The second major challenge is that GPUs come with faster but smaller physical memories than CPUs.
The limited GPU memory capacity can severely limit parallelism because there may not be sufficient memory for tracking the execution state of the large number of threads that traverse search trees and subtrees in parallel. 
While there has been work on solving related problems on GPUs, such as finding maximal cliques~\cite{mcg1, lessons, mcg3,bitsets,mcg4,mcg5} and generalized sub-graph matching~\cite{pangolin, sub0, sub5}, little attention has been given to $k$-clique counting in particular.
To the best of our knowledge, there are no performant parallel solutions specialized for $k$-clique counting on GPUs.

\begin{figure*}[t]
\centering
    \includegraphics[width=\textwidth]{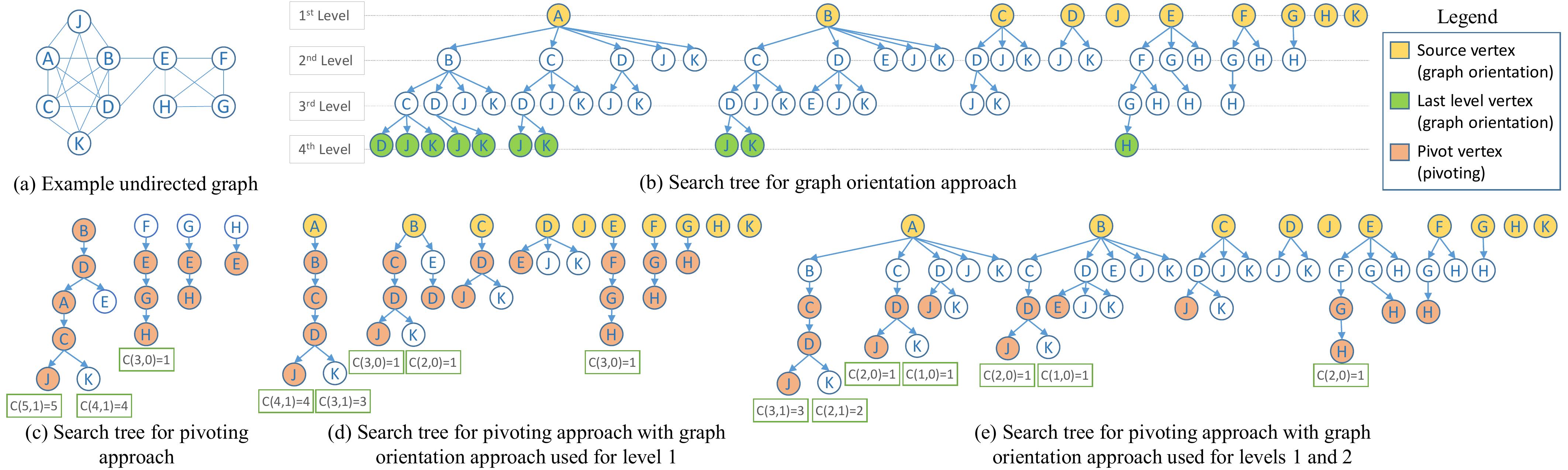}
    \caption{Counting 4-cliques in an example graph using different approaches}\label{fig:orientation-vs-pivot}
\end{figure*}

In this paper, we propose a novel parallel GPU solution to the $k$-clique counting problem.
Our proposed solution supports both graph orientation and pivoting for eliminating redundant clique discovery.
It incorporates both vertex-centric and edge-centric parallelization schemes for distributing work across thread blocks, as well as different ways for partitioning work within each thread block to extract fine-grain multi-level parallelism while tolerating load imbalance.
It uses binary encoding of induced sub-graphs to drastically reduce memory consumption while allowing for highly parallel list intersection operations.
It also takes advantage of the new independent thread scheduling support in recent GPUs (Volta and beyond) to allow threads to collaborate at sub-warp granularity for more effective multi-level parallelization and better utilization of parallel execution resources.
It further employs various other techniques to limit memory consumption.
Finding the right combination of optimizations that work together effectively is a key contribution of our work, and we believe that the insights from our optimization flow can be useful for optimizing other clique finding and graph mining solutions on GPUs.

Our evaluation shows that our best GPU implementation significantly outperforms the best state-of-the-art parallel CPU implementation by a geometric mean of 12.39$\times$, 6.21$\times$, and 18.99$\times$ for $k=4$, $7$, and $10$, respectively.
Our parallel solution scales to graphs with billions of edges for arbitrary values of $k$. 
We perform a detailed evaluation of the trade-offs between the vertex-centric and the edge-centric parallelization schemes, particularly pertaining to their impact on load imbalance and their interaction with the two redundancy elimination approaches and different values of $k$.
We also show that binary encoding improves performance by a 2.17$\times$ and 1.38$\times$ and that sub-warp partitioning improves performance by 1.98$\times$ and 1.73$\times$ for graph orientation and pivoting, respectively.

%% file: sec/2-background.tex
\section{Background}

\subsection{Clique Counting}

A common approach to counting $k$-cliques in a graph is to traverse a search tree for each vertex in the graph to find $k$-cliques that contain that vertex.
The search tree for each vertex (1-clique) branches out to the vertex's neighbors to find edges (2-cliques), then for each edge, branches out to the common neighbors of the edge's end-points to find triangles (3-cliques), then for each triangle, branches out to the common neighbors of the vertices in the triangle to find 4-cliques, and so on.
In general, for each $(k-1)$-clique, the tree branches out to the common neighbors of the $k-1$ vertices in the clique to find the $k$-cliques.
This approach to $k$-clique counting is commonly parallelized by processing different trees or subtrees in parallel.

One key distinguishing feature among algorithms is how they avoid discovering the same clique redundantly from multiple root vertices.
Avoiding redundant clique discovery results in a substantial reduction in the amount of work done, thereby improving performance.
The two major approaches to avoiding redundant clique discovery are: (1) orienting the graph before traversal, and (2) pivoting.
These two approaches are described in Sections~\ref{sec:orient} and~\ref{sec:pivot}, respectively.

\subsection{Graph Orientation Approach}\label{sec:orient}

Graph orientation (or vertex ordering)  is a preprocessing step that transforms the graph from an undirected graph to a directed one. 
Common orientation criteria include vertex degree~\cite{chiba1985arboricity, arb-count, degreeOrdering1}, graph coloring~\cite{orderingheuristic}, and degeneracy based on $k$-core decomposition~\cite{bkpivoting, kclist, arb-count} or relaxations of $k$-core~\cite{arb-count}.
Graph orientation relies on the fact that a clique is a symmetric substructure, hence, it can be found by starting from any vertex it contains.

Fig.~\ref{fig:orientation-vs-pivot}(b) shows how the graph in Fig.~\ref{fig:orientation-vs-pivot}(a) is explored in the graph orientation approach to find all the 4-cliques.
Assume that the edges are oriented in alphabetical order (from the earlier letter to the later letter).
The first level contains all the vertices in the graph representing the root of their respective search trees.
At the second level, each tree branches out from the root vertex to its neighbors.
For example, the branch $A \rightarrow B$ indicates that there is an edge from vertex $A$ to vertex $B$.
At the third level, each edge branches out to the triangles it participates in.
For example, the path $A \rightarrow B \rightarrow C$ indicates that there is a triangle containing vertices $A$, $B$, and $C$.
Here, C is found by intersecting the adjacency lists of vertices $A$ and $B$ (i.e., $Adj(A) \cap  Adj(B)$, where $Adj(v)$ is the adjacency list of a vertex $v$).
Finally, at the fourth level, each triangle branches out to the 4-cliques it participates in.
For example, the path $A \rightarrow B \rightarrow C \rightarrow D$ indicates that there is a 4-clique containing vertices $A$, $B$, $C$, and $D$.
Here, $D$ is found by intersecting the adjacency lists for $A$, $B$, and $C$.
Since, $Adj(A) \cap Adj(B)$ was computed in the previous level, what remains is intersecting the previous result with $Adj(C)$.
Since we are looking for 4-cliques, the tree traversal stops at the fourth level.
In general, when looking for $k$-cliques, the tree traversal stops at level $k$.

Graph orientation has two main benefits.
The first benefit is that it eliminates redundant clique discovery as previously mentioned.
In the example in Fig.~\ref{fig:orientation-vs-pivot}(b), the 4-clique containing vertices $A$, $B$, $C$, and $D$ is only discovered in the tree rooted at vertex $A$.
It is not redundantly discovered in the other trees because vertex $A$ is not reachable from the other vertices in the directed graph.
The second benefit of graph orientation is that it reduces the out-degrees of the vertices and the maximum out-degree of the graph.
The maximum out-degree has a quadratic impact on memory consumption, as we show in Section~\ref{sec:ind}.

\begin{figure}
    \small
    \centering
    \input{fig/2-background/graph-orient-recursive}
    \caption{Tree Traversal for Graph Orientation}\label{fig:graph-orientation-recursive}
    \vspace{-10pt}
\end{figure}

Fig.~\ref{fig:graph-orientation-recursive} shows the pseudocode for traversing a subtree in the graph orientation approach.
The parameters are the graph $G$, the value of $k$, the current level $\ell$, and a set of vertices $I$.
$I$ contains the intersection of the adjacency lists of all the vertices in the previous levels which is also the set of vertices at the current level representing the set of cliques at the current level.
For example, if we are visiting level 3 along the path  $A \rightarrow B$, then $\ell$ will be 3 and $I$ will be $Adj(A) \cap  Adj(B)$ representing the common neighbors of $A$ and $B$ or the triangles that $A$ and $B$ are part of.
The pseudocode iterates through the vertices at the current level ($I$) which represent the $\ell$-cliques (line 3).
For each vertex $v$, it intersects that vertex's adjacency list $Adj_G(v)$ with those of the previous levels' vertices ($I$) to find the vertices at the next level ($I'$) which represent the $(\ell + 1)$-cliques (line 4).
If $\ell + 1$ is $k$ (line 5), then the total number of $k$-cliques is incremented by the number of $k$-cliques that were just found (line 6).
Otherwise, if the set of $(\ell + 1)$-cliques found is not empty (line 7), then they are visited at the next level (line 8).

Parallel $k$-clique counting based on graph orientation has been extensively studied on CPUs~\cite{arb-count, kclist, orderingheuristic, chiba1985arboricity}.
Vertices or edges are typically distributed across CPU threads, and each thread performs a sequential depth-first traversal on the trees or subtrees corresponding to the vertices or edges assigned to the thread to count the underlying cliques.
To the best of our knowledge, ARB-COUNT~\cite{arb-count} is the best performing parallel CPU implementation of $k$-clique counting based on graph orientation.
In this paper, we present our parallel GPU implementation of $k$-clique counting based on graph orientation and compare its performance with ARB-COUNT.

\subsection{Pivoting Approach}\label{sec:pivot}

Another approach to avoiding redundant clique discovery is pivoting~\cite{powerpivoting}.
The idea of pivoting is inspired by the Bron-Kerbosch maximal clique finding approach~\cite{bkpivoting}.
Pivoting relies on the fact that a $(k+i)$-clique consists of ${k+i} \choose {k}$ $k$-cliques, so instead of searching for all of these $k$-cliques, it is sufficient to find the largest $(k+i)$-clique and all the $k$-cliques it contains are found.
For example, the graph in Fig.~\ref{fig:orientation-vs-pivot}(a) contains a 5-clique consisting of vertices $A$, $B$, $C$, $D$, and $J$.
This 5-clique contains ${5 \choose 4} = 5$ different 4-cliques.
In the graph oriented approach in Fig.~\ref{fig:orientation-vs-pivot}(b), these five 4-cliques are discovered by five different paths in the search trees.
Instead, the pivoting approach just discovers the 5-clique and then concludes the existence of five 4-cliques.

Fig.~\ref{fig:orientation-vs-pivot}(c) shows an example of how the graph in Fig.~\ref{fig:orientation-vs-pivot}(a) is explored using the pivoting approach.
As the search tree is traversed, at every branching point in the search tree, one pivot child vertex is selected which is typically the vertex that has the largest common number of neighbors with its parent.
All the pivot's neighbors are then excluded while branching to the next level since these neighbors are eventually reachable from the pivot vertex (i.e., the pivot vertex is their parent in the search tree).
For example, at the first level in Fig.~\ref{fig:orientation-vs-pivot}(c), vertex $B$ is selected as the pivot vertex because it has the largest number of neighbors.
Accordingly, all of $B$'s neighbors ($A$, $C$, $D$, $E$, $J$, $K$) are excluded from creating search trees at the first level.
At the second level, when branching from vertex $B$ to its neighbors, vertex $D$ is selected as the pivot because it has the largest number of common neighbors with $B$.
Accordingly, all of $B$ and $D$'s common neighbors ($A$, $C$, $E$, $J$, $K$) are excluded while branching to the second level.
The tree traversal proceeds in this way.
Unlike the graph orientation approach, the pivoting approach does not stop at level $k$ because it is searching for the largest $(k+i)$-clique.
It continues exploring until it reaches the bottom of the tree (or satisfies a stopping criteria~\cite{powerpivoting}).
To calculate the number of $k$-cliques found on a path, the combinatorial formula $n_p \choose n_v - k$ is used, where $n_p$ is the number of pivots in the path, and $n_v$ is the number of vertices in the path.

\begin{figure}
    \small
    \centering
    \input{fig/2-background/pivoting-recursive}
    \caption{Tree Traversal for Pivoting}\label{fig:pivoting-recursive}
    \vspace{-10pt}
\end{figure}

Fig.~\ref{fig:pivoting-recursive} shows the pseudocode for traversing a subtree in the pivoting approach.
Compared to Fig.~\ref{fig:graph-orientation-recursive}, it takes an additional parameter to track the number of pivot vertices encountered on the path.
First, the pivot vertex is found (line 3) and the neighbors of the pivot vertex are pruned from the level (line 4).
Next, the code iterates through the remaining vertices which represent the $\ell$-cliques (line 5).
If the stopping criteria~\cite{powerpivoting} has not been reached (line 7), then the vertex's adjacency list $Adj_G(v)$ is intersected with those of the previous levels' vertices ($I$) to find the vertices at the next level ($I'$) which represent the $(\ell + 1)$-cliques (line 8).
$I'$ also excludes vertices at the current level that have already been visited to avoid finding redundant cliques (line 8).
If $I'$ is not empty (line 9) meaning that some $(\ell + 1)$-cliques are found, then the next level is visited (line 10).
Otherwise, if there are no $(\ell + 1)$-cliques, then the current tree node represents the largest clique on this path.
If the size of this large clique is $\ge k$ (line 11), then the total number of $k$-cliques is incremented by the number of $k$-cliques in the large clique just found (line 12).

Compared to graph orientation, pivoting has the advantage that it reduces the search space significantly by eliminating the neighbors of the pivot vertex from the branching.
This reduction is clear when comparing Fig.~\ref{fig:orientation-vs-pivot}(b) and Fig.~\ref{fig:orientation-vs-pivot}(c).
The reduction in branching makes pivoting particularly suitable for large $k$, or for counting all cliques for all $k$.
On the other hand, pivoting has several disadvantages.
The first disadvantage of pivoting is that it requires deeper exploration of the search tree which exacerbates load imbalance in parallel implementations.
The second disadvantage of pivoting is that it reduces the amount of parallelism available by eliminating some of the search trees or subtrees and folding them into fewer and deeper trees.
The third disadvantage of pivoting is that it requires an undirected graph, which makes adjacency list intersection operations prohibitively expensive for large graphs where vertices have very high degrees, hence very long adjacency lists.
The second and third disadvantages are mitigated~\cite{powerpivoting} by starting with the graph orientation approach for the first level of the tree using a directed graph, then switching to the pivoting approach for the remaining levels using an undirected induced sub-graph (see Section~\ref{sec:ind}).
Examples of this hybrid approach are illustrated in Fig.~\ref{fig:orientation-vs-pivot}(d) and Fig.~\ref{fig:orientation-vs-pivot}(e).

To our knowledge, Pivoter~\cite{powerpivoting} is the only implementation of $k$-clique counting based on pivoting.
The implementation is parallelized on the CPU by distributing vertices or edges across CPU threads, and having each thread perform a sequential depth-first traversal of the tree or subtree.
In this paper, we present our parallel GPU implementation of $k$-clique counting based on pivoting and compare its performance to Pivoter.

\subsection{Induced Sub-graph Optimization}\label{sec:ind}

As shown in Fig.~\ref{fig:graph-orientation-recursive} and Fig.~\ref{fig:pivoting-recursive}, both the graph orientation and pivoting approaches spend a significant amount of time performing adjacency list intersection operations.
Note that set difference can also be performed as an intersection operation because $A - B = A \cap \overline{B}$.
A common optimization for the intersection operations is to shrink the size of the adjacency lists by removing from the graph, for each search tree, the vertices that will never be reached by the tree.
For example, in Fig.~\ref{fig:orientation-vs-pivot}, when traversing a tree rooted at the vertex $A$, only the neighbors of $A$ can ever be reached.
Hence, any vertex that is not a neighbor of $A$ can be removed from the graph before the traversal.

In general, when traversing a search tree rooted at a vertex $v$, the first step is to extract the vertex-induced sub-graph consisting of the vertices in $Adj(v)$.
This induced sub-graph is used throughout the tree traversal instead of the full graph.
Since the induced sub-graph is typically much smaller than the full graph, it has smaller adjacency lists resulting in faster adjacency list intersection operations.
Note that in principle, an induced sub-graph may be extracted at any level in the search tree.
For example, if the tree contains a path $v_1 \rightarrow v_2 \rightarrow ... \rightarrow v_i$, the subtree rooted at $v_i$ can only reach vertices that are neighbors of all the vertices $v_1, v_2, ..., v_i$.
Therefore, the induced sub-graph consisting of the vertices in $Adj(v_1) \cap Adj(v_2) \cap ... \cap Adj(v_i)$ is sufficient for traversing the subtree.
However, extracting the induced sub-graph at each level is usually not worth the overhead.
The induced sub-graph is typically extracted at one level, either the first or the second.
In this paper, both alternatives are explored.

The largest possible induced sub-graph has $d_{max}$ vertices, where $d_{max}$ is the maximum out-degree of the graph.
Therefore, the upper bound on the size of an induced sub-graph is $O(d_{max}^2)$.
Since the maximum out-degree of the graph has a quadratic impact on memory consumption, the choice of graph orientation is critical for reducing memory consumption, as mentioned in Section~\ref{sec:orient}.
It becomes even more critical in parallel implementations because when the trees or subtrees are processed in parallel and each has a different induced sub-graph, each needs its own memory space to store its induced sub-graph.
Section~\ref{sec:ind-bin} describes how the memory consumption of the induced sub-graphs is further reduced in our parallel GPU implementation.

%% file: fig/2-background/graph-orient-recursive.tex
\hrulefill
\begin{pseudo}
$numCliques = 0$ \\
\textbf{procedure} $traverseSubtree(G, k, \ell, I)$ \\+
    \textbf{for} $v \in I$ \\+
        $I' = I \cap Adj_G(v)$ \\
        \textbf{if} $\ell + 1 == k$ \\+
            $numCliques~+= |I'|$ \\-
        \textbf{else if} $|I'| > 0 $ \\+
            $traverseSubtree(G, k, \ell + 1, I')$ \\-
\end{pseudo}
\hrulefill

%% file: fig/2-background/pivoting-recursive.tex
\hrulefill
\begin{pseudo}
$numCliques = 0$ \\
\textbf{procedure} $traverseSubtree(G, k, \ell, I, nPivots)$ \\+
    $v_{pivot} = findPivotVertex(I, G)$ \\
    $I_{pruned} = I - Adj_G(v_{pivot})$ \\
    \textbf{for} $v \in I_{pruned}$ \\+
        $nPivots' = (v == v_{pivot})?(nPivots + 1):nPivots$ \\
        \textbf{if} $\ell + 1 - k \leq nPivots'$ \\+
            $I' = I \cap Adj_G(v) - \{u \in I_{pruned} | u < v\}$ \\
            \textbf{if} $|I'| > 0$ \\+
                $traverseSubtree(G, k, \ell + 1, I', nPivots')$ \\-
            \textbf{elseif} $\ell + 1 \geq k$ \\+
                $numCliques~+= {{nPivots'} \choose {\ell + 1 - k}}$
\end{pseudo}
\hrulefill

%% file: sec/3-implementation.tex
\section{Parallel Clique Counting on GPUs}

\subsection{Graph Format and Orientation Criteria}\label{sec:format}

We represent the input graph using the hybrid Compressed Sparse Row (CSR) + Coordinate (COO) storage format.
The CSR representation facilitates finding the adjacency list of a given vertex, which is useful for vertex-centric processing and parallelization.
The COO representation facilitates finding the source and destination vertex of a given edge, which is useful for edge-centric processing and parallelization.

Before clique counting begins, we first orient the graph to become a directed graph.
Recall that both the graph orientation approach and the pivoting approach require the graph to be oriented at the beginning.
Our implementation supports two different orientation criteria: degree orientation and degeneracy orientation.
Degree orientation orients edges from the vertex with the lower degree to the vertex with the higher degree.
Degeneracy orientation orients edges from the lower $k$-core order to the higher $k$-core order.
The $k$-core order is obtained from $k$-core decomposition which iteratively eliminates the minimum degree vertices from the graph.
The $k$-core order is the order in which the vertex is removed from the graph.

We implement both orientation criteria on the GPU.
In both cases, after determining which edges to keep, the undesired edges are filtered out and the CSR row pointers are recomputed with histogram and exclusive scan operations.
The tradeoff between orientation criteria is evaluated in Section~\ref{sec:eval-orientation}.

We note that although degree orientation and degeneracy orientation are currently supported, our implementation can easily be extended to support other orientation criteria.
The choice of orientation criteria is orthogonal to our work and not intended as a contribution of this paper.

\subsection{Parallelization Schemes}\label{sec:par}

GPUs provide a massive amount of parallelism and are capable of running tens of thousands to hundreds of thousands of threads concurrently~\cite{hwu2022programming}.
A grid of threads running on a GPU is divided into thread blocks.
Threads in the same thread block can collaborate by synchronizing at a barrier and sharing a fast scratchpad memory (also called shared memory).
Thread blocks are divided into warps which consist of 32 threads bound by the SIMD execution model.
Threads in the same warp can collaborate using low cost warp-level primitives.

Our main strategy for parallelizing $k$-clique counting on GPUs is to traverse different search trees or subtrees in parallel.
For both graph orientation and pivoting, we implement two different parallelization schemes: a vertex-centric scheme and an edge-centric scheme\footnote{The terms \textit{vertex-centric} and \textit{edge-centric} in the context of parallel graph processing commonly refer to when different parallel workers are assigned to different vertices or edges, respectively. In some parts of the literature, they also imply that a parallel worker is restricted to accessing only its vertex's incident edges or its edge's endpoints, respectively. In our usage of the terms, we do not assume this restriction.}.
In the vertex-centric scheme, each thread block is assigned to a vertex of the input graph (level 1 in the search tree) and is responsible for traversing the tree rooted at that vertex.
The threads in the block collaborate to extract the induced sub-graph consisting of the vertex's neighbors, then proceed to traverse the vertex's tree using the induced sub-graph.
In the edge-centric scheme, each thread block is assigned to an edge of the input graph (level 2 in the search tree) and is responsible for traversing the subtree stemming from that edge.
The threads in the block collaborate to extract the induced sub-graph consisting of the common neighbors of the edge's endpoints, then proceed to traverse the edge's subtree using the induced sub-graph.

The advantage of the vertex-centric scheme over the edge-centric scheme is that it extracts an induced sub-graph for each vertex's tree as opposed to each edge's subtree.
Hence, the vertex-centric scheme amortizes the cost of extracting the induced sub-graph over the traversal of a larger tree.
The advantage of the edge-centric scheme over the vertex-centric scheme is that it extracts more induced sub-graphs that are smaller in size.
Hence, the edge-centric scheme exposes finer-grain parallelism which makes it more robust against load imbalance.
It also results in shorter adjacency lists.
We compare the performance of the vertex-centric and edge-centric schemes in Section~\ref{sec:eval-opt}.

\begin{figure}[t]
\centering
    \includegraphics[width=0.9\columnwidth]{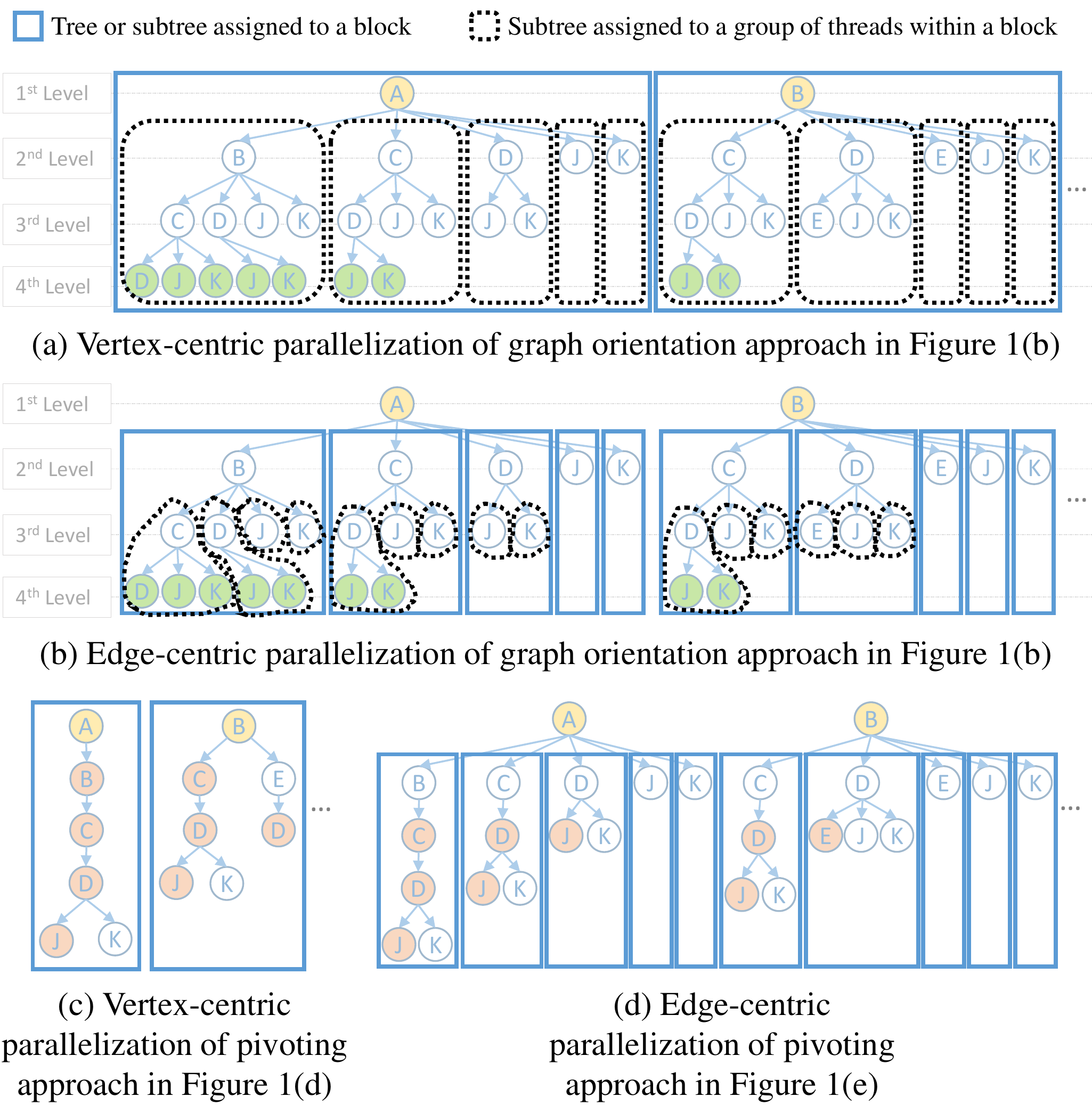}
    \caption{Parallelization schemes}\label{fig:par}
\end{figure}

Parallelization of work within each thread block differs between the graph orientation approach and the pivoting approach.
In the graph orientation approach, we partition the blocks into groups of threads and each group independently traverses one of the subtrees in the next level.
In the vertex-centric scheme, each group of threads is assigned to an outgoing edge (level 2 in the search tree) of the block's vertex, and the group independently traverses the subtree rooted at the edge.
An example is shown in Fig.~\ref{fig:par}(a).
In the edge-centric scheme, each group of threads is assigned to a triangle (level 3 in the search tree) that the block's edge participates in, and the group independently traverses the subtree rooted at that triangle.
An example is shown in Fig.~\ref{fig:par}(b).
In both cases, the threads in a group jointly perform a depth-first traversal of the subtree that the group is assigned to, visiting the nodes in the subtree sequentially.
At each node of the subtree, the threads in the group collaborate to perform the adjacency list intersection operation in parallel.
We discuss how we parallelize the adjacency list intersection operations in Section~\ref{sec:ind-bin}.

In the pivoting approach, we also partition the blocks into groups of threads, however these groups do not process next-level subtrees independently.
Instead, all threads in the block stay together as they jointly perform a sequential depth-first traversal of the tree/subtree that the block is assigned to.
However, at each node in the tree, identifying which neighbor is the pivot vertex (line 3 in Fig.~\ref{fig:pivoting-recursive}) requires performing a list intersection operation for each of the neighbors.
Therefore, each group of threads is assigned to a different neighbor and performs a list intersection operation for that neighbor to check if it is the pivot.
Examples of the vertex-centric and edge-centric schemes for the pivoting approach are shown in Fig.~\ref{fig:par}(c) and Fig.~\ref{fig:par}(d), respectively.

There are two reasons why, in the pivoting approach, and unlike the graph orientation approach, groups of threads are not assigned to process the next-level subtrees independently.
The first reason is that the process of finding the pivot element at each tree node is expensive and already provides enough work to be parallelized across groups.
The second reason is that the pivoting approach has deeper trees than the graph orientation approach, thereby requiring more memory to store intermediate results.
Parallelizing the next-level subtrees across groups of threads would require too much memory for storing the intermediate results of each subtree.

For now, a group of threads can be thought of as a warp, which is the most natural way to partition a block.
However, we improve on this partitioning granularity in Section~\ref{sec:sub-warp}.

\subsection{Traversing a Subtree}\label{sec:iterative-trav}

In Section~\ref{sec:par}, we saw how trees or subtrees are distributed across blocks or groups of threads to be traversed in parallel.
Tree traversal on CPUs is typically done using recursion.
However, using recursion on the GPU is not suitable because there is a large amount of intermediate traversal data that needs to be saved, and the tree is traversed jointly by multiple fine-grain threads that need to access the same intermediate traversal data.
For this reason, our implementation replaces recursion with an iterative tree traversal whereby threads traversing the same tree explicitly manage a shared stack.
We omit the details of how the recursive traversal is replaced with an iterative traversal due to space constraints.

The shared stack used in the iterative traversal is pre-allocated and provisioned for the maximum depth of the tree.
The large components of each stack entry (such as vertex arrays) are preallocated in global memory, while the small components (such as counters) are pre-allocated in shared memory for fast access.
A different stack is needed for each block or group that traverses a tree or subtree independently, which puts high pressure on the global memory capacity.
To reduce this pressure, we employ various memory management techniques discussed in Section~\ref{sec:mem-manage}.

\subsection {Binary Encoding of Induced Sub-graphs} \label{sec:ind-bin}

Recall from Section~\ref{sec:ind} that the first step a thread block performs before traversing its tree or subtree is to extract an induced sub-graph.
In the vertex-centric scheme, the induced sub-graph consists of the vertex's neighbors, whereas in the edge-centric scheme, the induced sub-graph consists of the common neighbors of the edge's endpoints.
We use a directed induced sub-graph for the graph orientation approach, and an undirected induced sub-graph for the pivoting approach.
The block's induced sub-graph is used by all threads in the block to perform the adjacency list intersection operations throughout the tree traversal.

The storage format used for the induced sub-graph is of utmost importance for both memory consumption and execution time.
For memory consumption, since each thread block has a different induced sub-graph, enough memory must be allocated for all the blocks running simultaneously so they can each store a private induced sub-graph.
Hence, the space efficiency of the induced sub-graph storage format is crucial for the overall memory consumption.
For execution time, since threads use the induced sub-graph to perform adjacency list intersection operations, the storage format must be designed to enable low-latency parallel intersections.

To optimize both memory consumption and execution time, we use binary encoding to represent the induced sub-graph.
To the best of our knowledge, our work is the first to use binary encoding for the induced sub-graphs in $k$-clique counting.
Related work on the maximal clique problem uses binary encoding for the entire graph~\cite{bitsets} or for specialized data structures to represent and operate on the candidate maximal cliques~\cite{mcg4, mcg5}.
Other graph processing works use binary encoding in different contexts.

\begin{figure}[t]
\centering
    \includegraphics[width=0.9\columnwidth]{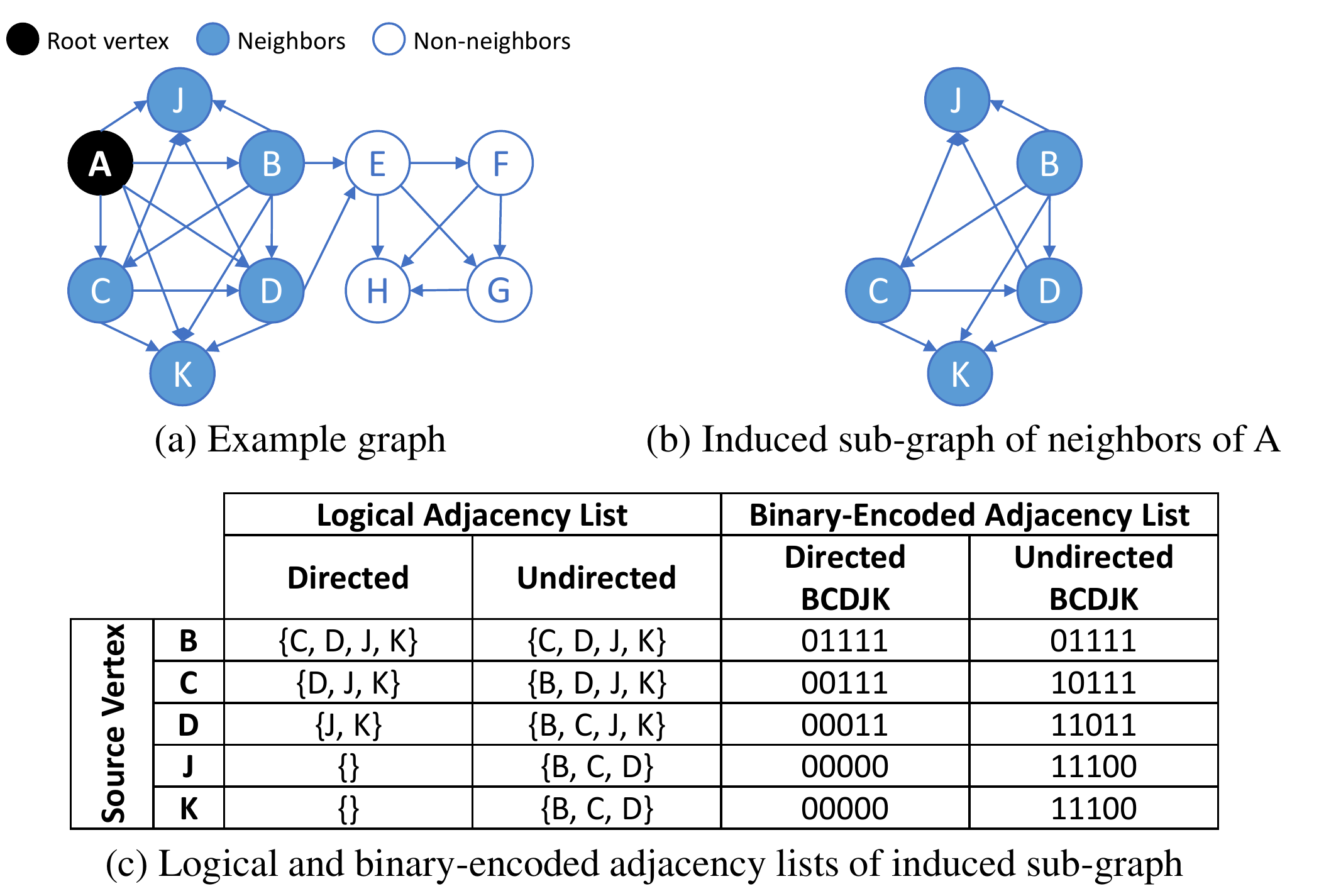}
    \caption{Binary encoding example}\label{fig:ind-bin}
\end{figure}

Fig.~\ref{fig:ind-bin} shows an example of how an induced sub-graph can be binary encoded.
Assume that a thread block is assigned to traverse the tree rooted at vertex A in the graph in Fig.~\ref{fig:ind-bin}(a).
The only vertices ever visited in that tree are the neighbors of A.
Therefore, the thread block starts by extracting the induced sub-graph consisting of the neighbors of A shown in Fig.~\ref{fig:ind-bin}(b).
This induced sub-graph is binary encoded as shown in Fig.~\ref{fig:ind-bin}(c).
The adjacency list of each vertex in the sub-graph consists of a bit vector with a 1 for each neighbor and a 0 otherwise. 
In this example, the neighbor vertices are assigned to bit positions in alphabetical order.
Any two lists can be intersected by performing a simple bitwise-and operation between the two bit vectors.
Note that in addition to the induced sub-graph being binary encoded, all the intermediate vertex lists (i.e., $I$, $I'$, and $I_{pruned}$ in Fig.~\ref{fig:graph-orientation-recursive} and Fig.~\ref{fig:pivoting-recursive}) are also binary encoded, and intersections with these intermediate vertex lists (line 4 in Fig.~\ref{fig:graph-orientation-recursive} and lines 4 and 8 in Fig.~\ref{fig:pivoting-recursive}) also use bitwise-and.

The advantage of binary encoding for memory consumption is that each vertex in an adjacency list is represented with a single bit.
Since dynamic memory allocation is not efficient on GPUs, the space for each thread block to store its induced sub-graph and its intermediate vertex lists must be pre-allocated with enough capacity for the largest possible sub-graph.
The largest possible sub-graph may have $d_{max}$ vertices, where $d_{max}$ is the maximum out-degree of the graph, and the sub-graph may be completely dense.
In this case, storing the sub-graph would require $O(d_{max}^2)$ memory, and storing the intermediate vertex lists would require $O(d_{max})$ memory per level.
We show in Section~\ref{sec:eval-orientation} that with a proper orientation criterion, the value of $d_{max}$ remains manageable even for very large graphs.
Nevetheless, binary encoding has the advantage of reducing the amount of memory needed for storing the sub-graph and the intermediate vertex lists by a factor of 32.

The advantage of binary encoding for execution time is that it allows list intersection operations to be performed using simple bitwise-and operations.
Traditional adjacency list intersection techniques on GPUs are complex to parallelize, suffer from control divergence, and exhibit uncoalesced memory access patterns.
On the other hand, performing bitwise-and on a bit vector is easy to parallelize across threads in a group or block, does not suffer from control divergence, and enables coalescing of memory accesses.

The reason binary encoding is particularly effective for the induced sub-graphs and intermediate vertex lists is that they typically consist of a small number of vertices, especially when a good graph orientation criteria is used.
Moreover, the induced sub-graphs are typically denser than the full graph.
In contrast, binary encoding for the full graph is impractical because the full graph has many more vertices and is usually much more sparse, resulting in many wasted 0 bits.
For this reason, we continue to represent the full graph using the hybrid CSR+COO format (see Section~\ref{sec:format}) and only represent the induced sub-graphs using binary encoding.
To extract the binary encoded induced sub-graph from the full hybrid CSR+COO graph, we intersect the adjacency lists of the hybrid CSR+COO graph using binary-search-based intersections.

We evaluate the performance improvement of using binary encoded induced sub-graphs in Section~\ref{sec:eval-be}.

\subsection{Sub-warp Partitioning}\label{sec:sub-warp}

In Section~\ref{sec:par}, we explained that for both the graph orientation approach and the pivoting approach, our implementation partitions thread blocks into groups of threads and distributes the block's work across these groups.
In the graph orientation approach, a block is assigned to a tree or subtree, and has each group of threads process one of the next level subtrees in parallel.
In the pivoting approach, a block is also assigned to a tree or subtree, but the groups of threads jointly iterate over the tree nodes sequentially.
However, for each tree node, when determining which child vertex is the pivot, each group of threads is used to check a different child in parallel.

One important design consideration is the granularity at which thread blocks are partitioned into groups.
The most natural granularity is the warp because threads in the same warp are bound together by SIMD and are able to collaborate using low cost warp-level primitives.
However, the introduction of binary encoding makes the warp granularity often too coarse.
Recall that threads within a group collaborate to perform a single list intersection operation in parallel.
With binary encoding, each thread can intersect 32 list elements simultaneously with a single bitwise-and operation.
Therefore, to fully utilize all 32 threads in the warp, the intersection needs to contain 1024 list elements.
We show in Section~\ref{sec:eval-orientation} that with proper graph orientation criteria, the maximum out-degree (and consequently, the largest binary encoded list size) is often much smaller than that.
Therefore, partitioning blocks at the warp granularity would lead to underutilization of parallel execution resources.

To address this issue, we implement sub-warp partitioning where thread blocks are partitioned to groups smaller than a warp.
Since the NVIDIA Volta architecture, the 
\textit{independent thread scheduling} has enabled fine-grain collaboration between a subset of threads within a warp.
We leverage this feature to enable the creation of thread groups that are 32, 16, 8, 4, 2, or 1 threads in size.
The number of threads per group is a tunable parameter and the same traversal code works for any group size.
We evaluate the performance improvement of using sub-warp partitioning in Section~\ref{sec:eval-sub-warp}.

\subsection{Memory Management} \label{sec:mem-manage}

The issue of memory consumption is exacerbated on the GPU compared to the CPU for two key reasons.
The first reason is that the capacity of the device memory on a typical GPU is much smaller than the capacity of main memory on a typical CPU.
The second reason is that GPUs are massively parallel processors so they perform much more work in parallel, thereby requiring much more intermediate data to be stored simultaneously.
For example, a CPU may run tens of threads at a time, so it only needs enough memory to maintain that many different sets of induced sub-graphs and intermediate vertex lists.
In contrast, a GPU may run hundreds to thousands of thread blocks at once, so it needs enough memory to maintain hundreds to thousands of different sets of induced sub-graphs and intermediate vertex lists.
Therefore, with a lower memory capacity and a higher demand for memory, it becomes critical to manage memory efficiently on the GPU.

We have discussed multiple techniques that we use for reducing memory consumption throughout this paper.
In Section~\ref{sec:par}, we discuss how an induced sub-graph is extracted once per thread block and shared by multiple groups of threads.
In Section~\ref{sec:ind-bin}, we discuss how binary encoding of sub-graphs and intermediate vertex lists reduces their memory requirement.
In this section, we discuss one more technique for reducing memory consumption.

Recall that memory needs to be pre-allocated for each block to store the induced sub-graph and intermediate vertex lists that it uses.
If we assign one vertex or edge to each block, we will launch many more blocks than the number that can execute simultaneously, which means that the pre-allocated memory spaces will not always be utilized.
To mitigate this inefficiency, we instead launch the maximum number of thread blocks that can run simultaneously and reuse these thread blocks to process multiple vertices or edges (by incrementing a global counter).
In doing so, we reduce the number of pre-allocated memory spaces and reuse the same memory space to process multiple vertices or edges.

%% file: sec/4-evaluation.tex
\section{Evaluation}\label{sec:eval}

\begin{table*}[t]
    \centering
    \caption{\change{Execution time and memory consumption of our GPU implementations}}\label{tab:compare}
    \resizebox{\textwidth}{!}{

\input{fig/4-evaluation/tab-results}
    }
\end{table*}

\subsection{Methodology}

\textbf{Evaluation Platform.} 
\change{In this section, } we evaluate our GPU implementations on an NVIDIA Volta V100 GPU with 32GB of memory.
The GPU is attached to an Intel Xeon Gold 6230 CPU.
We use a single CPU thread to drive the GPU.
\change{For a broader evaluation, we also report the execution times of our GPU implementations on an NVIDIA Ampere A100 GPU and an NVIDIA Ampere RTX 3090 GPU in Table~\ref{tab:compare_all} at the end of this paper.}

\textbf{Datasets.}
We use the same graphs used by ARB-COUNT~\cite{arb-count} for exact $k$-clique counting evaluation.
\change{The details of these graphs are shown in Table~\ref{tab:compare}.}
These graphs are real-world undirected graphs from the Stanford Network Analysis Project (SNAP)~\cite{snap}.

\textbf{Baselines.}
We compare our GPU implementations to two CPU baselines: ARB-COUNT~\cite{arb-count} and Pivoter~\cite{powerpivoting}.
ARB-COUNT~\cite{arb-count} represents the state-of-the-art parallel graph orientation implementation on CPU, which significantly outperforms other graph orientation implementations~\cite{kclist, orderingheuristic}.
Pivoter~\cite{powerpivoting} represents the state-of-the-art parallel pivoting implementation on CPU.
\change{In this section, } we use the execution times reported by ARB-COUNT~\cite{arb-count} for the parallel implementations of ARB-COUNT and Pivoter.
These times are obtained using an Intel Xeon Scalable (Cascade Lake) processor with 30 cores (\change{60 threads}) and 240~GB of main memory.
\change{For a broader evaluation, we also compare the execution times of our GPU implementations to the execution times reported by Lonkar and Beamer~\cite{lonkar2021accelerating} in Table~\ref{tab:compare_all} at the end of this paper.
These times are obtained using an Intel Xeon Platinum 8260 processor with 48 cores (96 threads) and 768~GB of main memory, but are only reported for up to $k=8$.
}

\textbf{Reporting of Measurements.}
The execution times we report include the time spent pre-processing and counting, and exclude the time spent \change{reading} the graph \change{from disk}.
Unless otherwise specified, we report the time achieved for the best combination of orientation criteria (degree or degeneracy), parallelization scheme (vertex-centric or edge-centric), and sub-warp partition size.
\change{However, we make suggestions for how to select a good combination of these parameters in Section~\ref{sec:eval-param-selection}.}
Similar to ARB-COUNT~\cite{arb-count}, we do not report times greater than five hours.

\begin{figure*}[t]
    \centering
    \includegraphics[width=\textwidth]{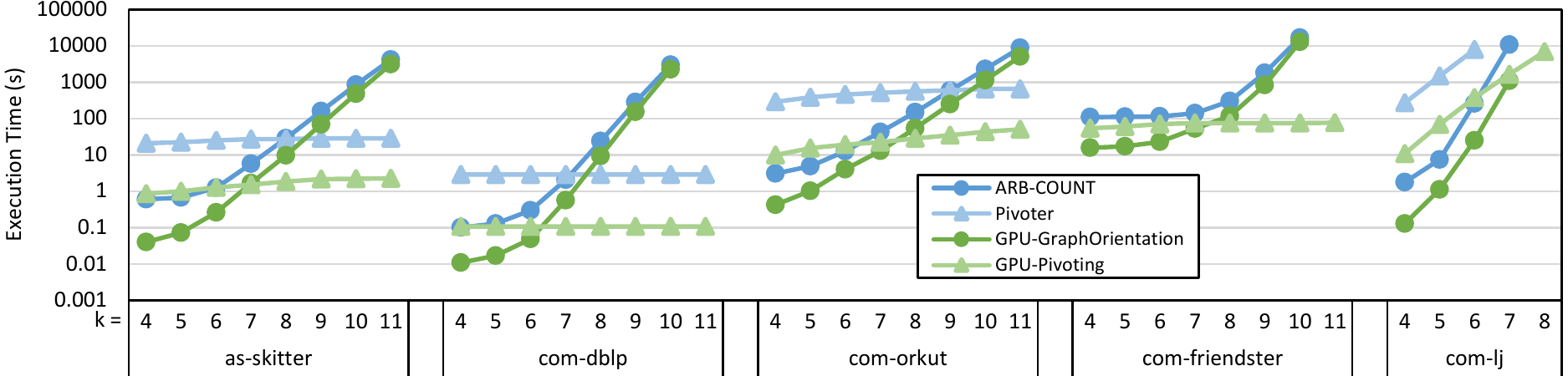}
    \caption{Comparing execution time with state-of-the-art parallel CPU implementations (lower is better)}\label{fig:baselines}
\end{figure*}

\begin{figure*}[t]
    \centering
    \includegraphics[width=\textwidth]{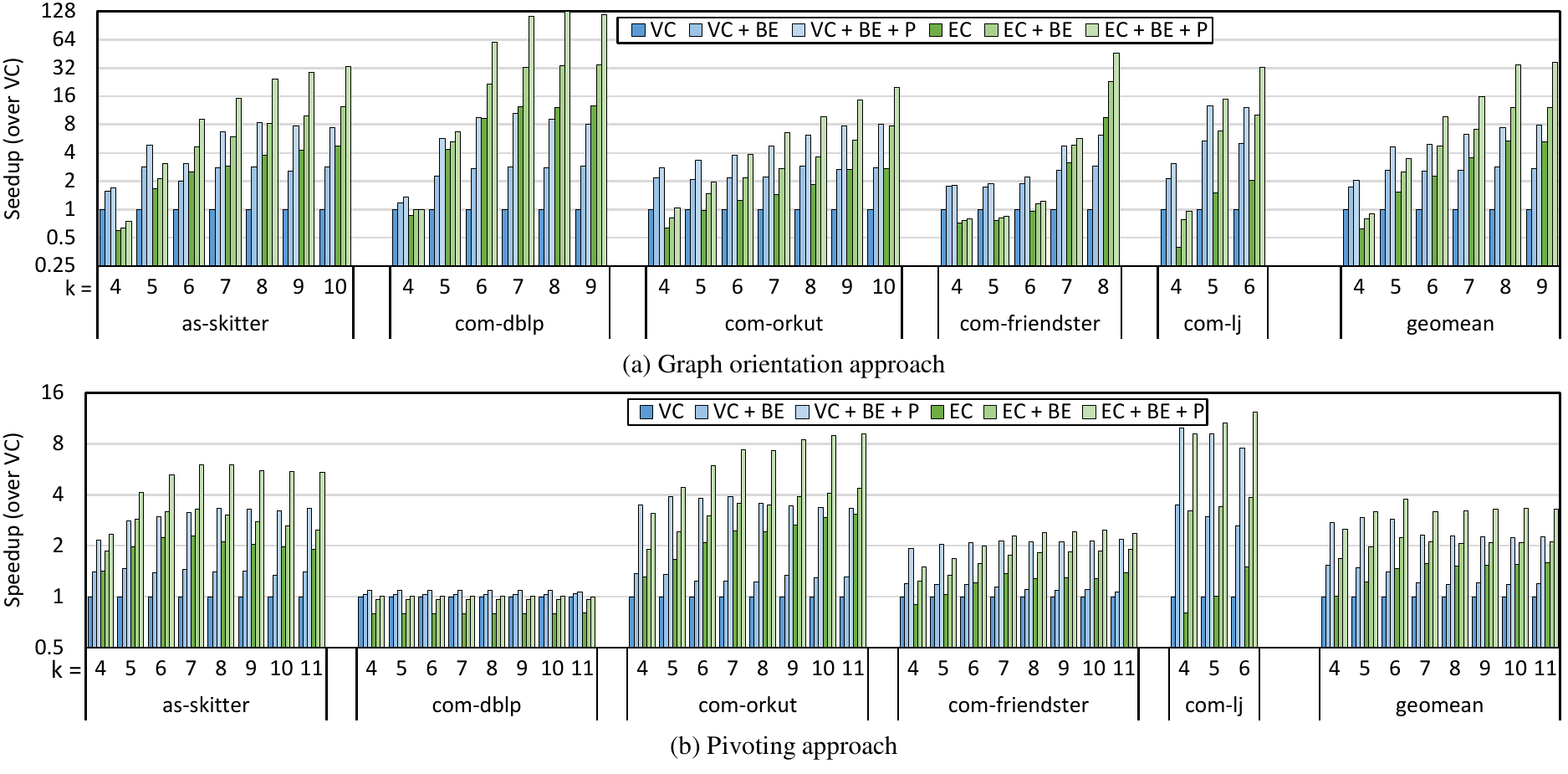}
    \caption{Impact of parallelization schemes and speedup of optimizations for each approach (higher is better)}\label{fig:opt}
\end{figure*}

\subsection{Comparison with Parallel CPU Implementations}\label{sec:eval-baselines}

Fig.~\ref{fig:baselines} compares the execution time of the state-of-the-art parallel CPU implementations with our GPU implementations for both the graph orientation approach and the pivoting approach.
\change{These execution times are also reported in Table~\ref{tab:compare}, along with details about each graph and the memory consumed by each of our implementations for each graph.}
The missing datapoints in the Fig.~\ref{fig:baselines} are those that take longer than 5~hrs to execute or run out of memory\change{, as shown in Table~\ref{tab:compare}}.
Note that our GPU implementations do not run out of memory for any scenario, despite the constrained GPU memory capacity.
Based on the results in Fig.~\ref{fig:baselines} \change{and Table~\ref{tab:compare}}, we make two key observations.

\textbf{Graph Orientation vs. Pivoting.}
The first observation is that the graph orientation approach performs better than the pivoting approach for small values of $k$, while the latter performs better for large values of $k$.
This observation is consistent with that made in prior work~\cite{arb-count}.
The observation applies for both CPU and GPU implementations.
Recall from Section~\ref{sec:pivot} that pivoting has the advantage of reducing the amount of branching but the disadvantage of having deeper search trees.
For small values of $k$, branching for the graph orientation approach is moderate, whereas the deep trees in the pivoting approach create load imbalance.
However, as $k$ gets larger, the branching drastically increases causing the graph orientation approach to suffer.
In most cases, the transition from the graph orientation approach being fastest to the pivoting approach being fastest happens at around $k=7$.

\textbf{GPU vs. CPU.}
The second observation is that our best GPU implementation consistently and significantly outperforms the best parallel state-of-the-art CPU implementation across all values of $k$.
Our best GPU implementation outperforms the best CPU implementation by a geometric mean speedup of 12.39$\times$, 6.21$\times$, and 18.99$\times$ for $k=4$, $7$, and $10$, respectively.
This result demonstrates the effectiveness of GPUs at accelerating $k$-clique counting.

\begin{figure*}
    \centering
    \includegraphics[width=\textwidth]{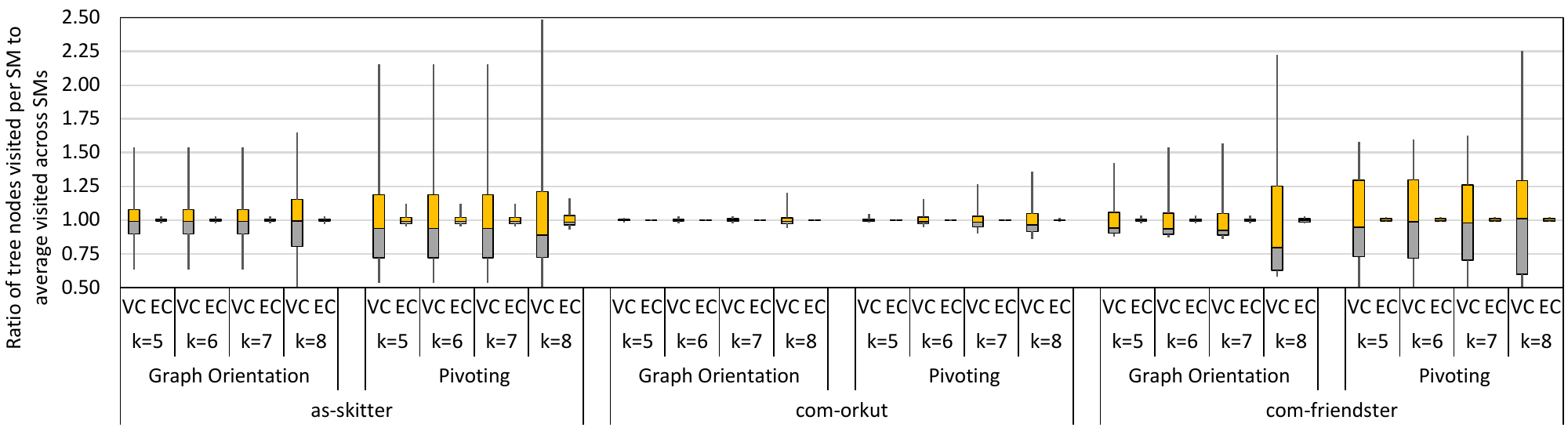}
    \caption{Load imbalance for a select set of graphs}\label{fig:load-balance}
\end{figure*}

\subsection{Impact of Parallelization Schemes and Optimizations}\label{sec:eval-opt}\label{sec:eval-be}\label{sec:eval-sub-warp}

Fig.~\ref{fig:opt}(a) and Fig.~\ref{fig:opt}(b) show the incremental speedup of binary encoding and sub-warp partitioning for both parallelization schemes for the graph orientation approach and the pivoting approach, respectively.
VC and EC refer to the vertex-centric and edge-centric parallelization schemes, respectively, with induced sub-graphs represented using the CSR format, parallel list intersections performed using the binary search approach, and blocks partitioned into groups at warp granularity.
The +BE suffix refers to when binary encoding is applied to the induced sub-graphs instead of using CSR and parallel list intersections are performed using bitwise-and operations.
The +P suffix refers when sub-warp partitioning is applied and the best partition size is used.
The omitted datapoints in the figure are those where the baseline (VC) takes longer than 5~hrs to run.

\textbf{Parallelization Scheme.}
We observe from Fig.~\ref{fig:opt} that the vertex-centric parallelization scheme is more effective than the edge-centric parallelization scheme for the initial values of $k$, particularly for the graph orientation approach.
For example, for $k=4$, VC+BE+P has a geometric mean speedup over EC+BE+P of 2.27$\times$ for the graph orientation approach and 1.09$\times$ for the pivoting approach.
However, as $k$ gets larger, the edge-centric parallelization scheme becomes more effective.
For example, for $k=9$, EC+BE+P has a geometric mean speedup over VC+BE+P of 4.68$\times$ for the graph orientation approach and 1.45$\times$ for the pivoting approach.
Recall from Section~\ref{sec:par} that the vertex-centric scheme has the advantage of amortizing the cost of extracting the induced sub-graph over a larger tree, whereas the edge-centric scheme has the advantage of extracting more parallelism making it more robust against load imbalance.
The graph orientation approach for small $k$ has the smallest trees, hence amortizing the induced sub-graph extraction across larger trees makes the vertex-centric scheme attractive.
However, as $k$ increases, load imbalance increases, which makes the edge-centric scheme attractive.
We also note that the transition from the vertex-centric scheme being fastest to the edge-centric scheme being fastest happens at a much smaller $k$ for our GPU implementation than it does for prior parallel CPU implementations~\cite{arb-count}.
GPUs exhibit more parallel execution resources than CPUs which makes them more sensitive to load imbalance, thereby favoring an earlier transition to the more load-balanced edge-centric scheme.
\change{Overall, we observe that selecting the vertex-centric scheme for $k < 6$ and the edge-centric scheme for $k \ge 6$ gives the best or near-best result in most cases.}

To further study the impact of the edge-centric scheme on load imbalance, Fig.~\ref{fig:load-balance} shows the distribution of work across SMs for a select set of graphs for different approaches and values of $k$.
The load of an SM is measured as the number of tree nodes visited by the SM normalized to the average number of tree nodes visited by all SMs.
As expected, the load imbalance is higher for pivoting than for graph orientation because pivoting has fewer and deeper search trees.
The load imbalance also increases with the value of $k$ because the search trees become deeper.
Most notably, we observe that the edge-centric scheme consistently has better load balance than the vertex-centric scheme.

\change{
For com-dblp with pivoting, we observe from Fig.~\ref{fig:opt}(b) that there is little performance impact from the choice of parallelization scheme, not to mention other optimizations.
The reason is that com-dblp has many clusters (it has the highest clustering coefficient~\cite{snap} among the graphs), which makes the pivoting optimization particularly effective on it, leaving little room for other optimizations to have a significant impact.
}

\textbf{Binary Encoding.}
We observe from Fig.~\ref{fig:opt} that binary encoding gives consistent performance improvement for both the graph orientation and pivoting approaches across all graphs, parallelization schemes, and values of $k$.
The geometric mean speedup of applying binary encoding is 2.17$\times$ for the graph orientation approach, and 1.38$\times$ for the pivoting approach.
Recall from Section~\ref{sec:ind-bin} that binary encoding improves execution time because it enables lower-latency parallel list intersection operations.
The graph orientation approach spends the majority of time performing list intersection operations, whereas the pivoting approach performs other kinds of operations like finding a maximum.
For this reason, it is expected that the speedup of binary encoding would be more pronounced in the graph orientation approach.

\textbf{Sub-warp Partitioning.}
We observe from Fig.~\ref{fig:opt} that sub-warp partitioning also gives consistent performance improvement for both the graph orientation and pivoting approaches across all graphs, parallelization schemes, and values of $k$.
The geometric mean speedup of applying sub-warp partitioning with the best partition size is 1.98$\times$ for the graph orientation approach, and 1.73$\times$ for the pivoting approach.
Recall from Section~\ref{sec:par} that the groups of threads within a block in the graph orientation approach operate completely independently, whereas in pivoting, these groups collaborate with each other to find the pivot vertex.
For this reason, it is expected that the speedup of unleashing more parallelism via sub-warp partitioning will be more pronounced in the graph orientation approach.

Regarding the choice of the best partition size, from our experience, partition sizes of 32 and 16 are never favorable.
Aside from these, selecting the wrong partition size results in a geometric mean reduction in speedup of 1.24$\times$ in the average case and 1.65$\times$ in the worst case, which is within the speedup margin of sub-warp partitioning.
Hence, sub-warp partitioning is still beneficial even if the best partition size is not correctly selected.
In addition, we have found that \change{for graph orientation,} graphs with lower maximum degree \change{(i.e., $< 200$)} favor having fewer threads per group \change{(e.g., 1 or 2)}, whereas graphs with higher maximum degree \change{(i.e., $\ge 200$)} favor having more threads per group (e.g., 8).
Graphs with higher maximum degree have larger intermediate adjacency lists that need to be interstected, so more threads can be utilized in performing the intersection operations in parallel.
\change{For pivoting, we have found that having fewer threads per group (e.g., 1 or 2) gives the best performance in most cases.}

\subsection{Impact of Graph Orientation Criteria}\label{sec:eval-orientation}

\begin{table}[t]
    \centering
    \caption{Impact of graph orientation criteria}\label{tab:orient}
    \vspace{-5pt}
    \resizebox{0.9\columnwidth}{!}{
        \input{fig/3-implementation/oritentation}
    }
\end{table}

Table~\ref{tab:orient} shows the achieved maximum out-degree of the graph and the pre-processing time of the two different orientation criteria used in this work, namely degree orientation and degeneracy orientation.
It is clear that degeneracy orientation achieves lower maximum out-degree but has higher pre-processing time.
Recall that the maximum out-degree forms an upper bound on the length of the adjacency list intersections.
Therefore, a lower maximum out-degree results in faster list intersection operations
throughout the traversal.

Our analysis of the best orientation criteria for different runs shows that \change{when the graph orientation approach is used,} runs with lower values of $k$ \change{(i.e., $k < 7$)} favor degree orientation, whereas runs with higher values of $k$ favor degeneracy orientation.
Runs with higher values of $k$ take longer and perform more list intersections, hence there is enough work reduction to amortize degeneracy orientation's higher pre-processing cost.
However, runs with lower values of $k$ do not perform enough list intersections to amortize the pre-processing cost.
\change{On the other hand, when pivoting is used, degeneracy orientation is always preferred.
Degeneracy orientation maximizes the effectiveness of pivoting at eliminating branches of the search tree.}

The trends observed in this subsection are consistent with the trends observed in prior work~\cite{arb-count}.
We include this analysis here for completeness, however, as mentioned in Section~\ref{sec:format}, the choice of orientation criteria is orthogonal to our work and not intended as a contribution.

\subsection{\change{Algorithm and Parameter Selection}}\label{sec:eval-param-selection}

\change{
Throughout this section, unless otherwise specified, we have reported results for the best choice of algorithm and optimization parameters to show the maximum potential of our approach.
However, when users solve for a particular graph and value of $k$, they face the challenge of selecting the best algorithm and optimization parameters.
In this subsection, we make recommendations for how to make this selection based on our empirical analysis.

There are four selections that need to be made in our approach:
    (1) the algorithm (graph orientation or pivoting),
    (2) the orientation criteria (degree or degeneracy),
    (3) the parallelization scheme (vertex-centric or edge-centric), and
    (4) the sub-warp partition size.
For selecting the algorithm, we observe in Section~\ref{sec:eval-baselines} that selecting graph orientation for $k<7$ and pivoting for $k \ge 7$ gives the best result in most cases.
For selecting the orientation criteria, we observe in Section~\ref{sec:eval-orientation} that graph orientation favors degree orientation when $k < 7$ and degeneracy orientation otherwise, whereas pivoting always favors degeneracy orientation.
For selecting the parallelization scheme, we observe in Section~\ref{sec:eval-opt} that selecting the vertex-centric scheme for $k < 6$ and the edge-centric scheme for $k \ge 6$ gives the best result in most cases.
For selecting the sub-warp partition size, we observe in Section~\ref{sec:eval-sub-warp} that for graph orientation, graphs with maximum degree $< 200$ favor partition sizes of 1 or 2, and graphs with maximum degree $\ge 200$ favor a partition size of 8, whereas for pivoting, a partition size of 1 is usually favorable.

Following these guidelines, users can select a near-optimal combination of algorithm and optimization parameters in the majority of cases.
Compared to when the best combination is selected every time, the execution times achieved if these guidelines are followed are only 1.17$\times$ slower (geometric mean), which is well within the margin of speedups reported in this paper.
}

\input{fig/4-evaluation/tab-extended}

%% file: fig/4-evaluation/tab-results.tex


\begin{tabular}{lcccr|rrrr|rr|}
\cline{6-11}
                                                      & \multicolumn{1}{l}{}                             & \multicolumn{1}{l}{}                                & \multicolumn{1}{l}{}    & \multicolumn{1}{c|}{}             & \multicolumn{4}{c|}{Execution Time (s)}                                                                                                                                                                       & \multicolumn{2}{c|}{Memory Consumption (MB)}                                \\ \hline
\multicolumn{1}{|c|}{Graph}                           & \multicolumn{1}{c|}{|V|}                         & \multicolumn{1}{c|}{|E|}                            & \multicolumn{1}{c|}{k}  & \multicolumn{1}{c|}{\# k-Cliques} & \multicolumn{1}{c|}{ARB-COUNT~\cite{arb-count}}            & \multicolumn{1}{c|}{Pivoter~\cite{powerpivoting}}              & \multicolumn{1}{c|}{\begin{tabular}[c]{@{}c@{}}GPU-Graph\\ Orientation\end{tabular}} & \multicolumn{1}{c|}{GPU-Pivot} & \multicolumn{1}{c|}{GPU-Graph Orientation} & \multicolumn{1}{c|}{GPU-Pivot} \\ \hline
\multicolumn{1}{|l|}{\multirow{8}{*}{as-skitter}}     & \multicolumn{1}{c|}{\multirow{8}{*}{1,696,415}}  & \multicolumn{1}{c|}{\multirow{8}{*}{11,095,298}}    & \multicolumn{1}{c|}{4}  & 148,834,439                       & \multicolumn{1}{r|}{0.60}                 & \multicolumn{1}{r|}{20.90}                & \multicolumn{1}{r|}{0.034}                                                           & 0.459                          & \multicolumn{1}{r|}{152.09}                & \multirow{8}{*}{149.21}        \\ \cline{4-10}
\multicolumn{1}{|l|}{}                                & \multicolumn{1}{c|}{}                            & \multicolumn{1}{c|}{}                               & \multicolumn{1}{c|}{5}  & 1,183,885,507                     & \multicolumn{1}{r|}{0.67}                 & \multicolumn{1}{r|}{22.54}                & \multicolumn{1}{r|}{0.069}                                                           & 0.721                          & \multicolumn{1}{r|}{154.59}                &                                \\ \cline{4-10}
\multicolumn{1}{|l|}{}                                & \multicolumn{1}{c|}{}                            & \multicolumn{1}{c|}{}                               & \multicolumn{1}{c|}{6}  & 9,759,000,981                     & \multicolumn{1}{r|}{1.24}                 & \multicolumn{1}{r|}{25.30}                & \multicolumn{1}{r|}{0.245}                                                           & 1.01                           & \multicolumn{1}{r|}{157.09}                &                                \\ \cline{4-10}
\multicolumn{1}{|l|}{}                                & \multicolumn{1}{c|}{}                            & \multicolumn{1}{c|}{}                               & \multicolumn{1}{c|}{7}  & 73,142,566,591                    & \multicolumn{1}{r|}{5.73}                 & \multicolumn{1}{r|}{27.14}                & \multicolumn{1}{r|}{1.434}                                                           & 1.269                          & \multicolumn{1}{r|}{152.09}                &                                \\ \cline{4-10}
\multicolumn{1}{|l|}{}                                & \multicolumn{1}{c|}{}                            & \multicolumn{1}{c|}{}                               & \multicolumn{1}{c|}{8}  & 481,576,204,696                   & \multicolumn{1}{r|}{28.38}                & \multicolumn{1}{r|}{27.46}                & \multicolumn{1}{r|}{9.531}                                                           & 1.585                          & \multicolumn{1}{r|}{162.09}                &                                \\ \cline{4-10}
\multicolumn{1}{|l|}{}                                & \multicolumn{1}{c|}{}                            & \multicolumn{1}{c|}{}                               & \multicolumn{1}{c|}{9}  & 2,781,731,674,867                 & \multicolumn{1}{r|}{158.45}               & \multicolumn{1}{r|}{28.45}                & \multicolumn{1}{r|}{68.221}                                                          & 1.835                          & \multicolumn{1}{r|}{164.59}                &                                \\ \cline{4-10}
\multicolumn{1}{|l|}{}                                & \multicolumn{1}{c|}{}                            & \multicolumn{1}{c|}{}                               & \multicolumn{1}{c|}{10} & 14,217,188,170,569                & \multicolumn{1}{r|}{854.87}               & \multicolumn{1}{r|}{28.45}                & \multicolumn{1}{r|}{474.785}                                                         & 1.777                          & \multicolumn{1}{r|}{167.09}                &                                \\ \cline{4-10}
\multicolumn{1}{|l|}{}                                & \multicolumn{1}{c|}{}                            & \multicolumn{1}{c|}{}                               & \multicolumn{1}{c|}{11} & 64,975,151,572,336                & \multicolumn{1}{r|}{4,158.53}             & \multicolumn{1}{r|}{28.45}                & \multicolumn{1}{r|}{2,997.385}                                                       & 1.777                          & \multicolumn{1}{r|}{169.59}                &                                \\ \hline
\multicolumn{1}{|l|}{\multirow{8}{*}{com-dblp}}       & \multicolumn{1}{c|}{\multirow{8}{*}{317,080}}    & \multicolumn{1}{c|}{\multirow{8}{*}{1,049,866}}     & \multicolumn{1}{c|}{4}  & 16,713,192                        & \multicolumn{1}{r|}{0.10}                 & \multicolumn{1}{r|}{2.88}                 & \multicolumn{1}{r|}{0.008}                                                           & 0.109                          & \multicolumn{1}{r|}{26.64}                 & \multirow{8}{*}{23.38}         \\ \cline{4-10}
\multicolumn{1}{|l|}{}                                & \multicolumn{1}{c|}{}                            & \multicolumn{1}{c|}{}                               & \multicolumn{1}{c|}{5}  & 262,663,639                       & \multicolumn{1}{r|}{0.13}                 & \multicolumn{1}{r|}{2.88}                 & \multicolumn{1}{r|}{0.016}                                                           & 0.109                          & \multicolumn{1}{r|}{29.14}                 &                                \\ \cline{4-10}
\multicolumn{1}{|l|}{}                                & \multicolumn{1}{c|}{}                            & \multicolumn{1}{c|}{}                               & \multicolumn{1}{c|}{6}  & 4,221,802,226                     & \multicolumn{1}{r|}{0.30}                 & \multicolumn{1}{r|}{2.88}                 & \multicolumn{1}{r|}{0.042}                                                           & 0.109                          & \multicolumn{1}{r|}{31.64}                 &                                \\ \cline{4-10}
\multicolumn{1}{|l|}{}                                & \multicolumn{1}{c|}{}                            & \multicolumn{1}{c|}{}                               & \multicolumn{1}{c|}{7}  & 60,913,718,813                    & \multicolumn{1}{r|}{2.05}                 & \multicolumn{1}{r|}{2.88}                 & \multicolumn{1}{r|}{0.545}                                                           & 0.109                          & \multicolumn{1}{r|}{26.64}                 &                                \\ \cline{4-10}
\multicolumn{1}{|l|}{}                                & \multicolumn{1}{c|}{}                            & \multicolumn{1}{c|}{}                               & \multicolumn{1}{c|}{8}  & 777,232,734,905                   & \multicolumn{1}{r|}{24.06}                & \multicolumn{1}{r|}{2.88}                 & \multicolumn{1}{r|}{9.031}                                                           & 0.109                          & \multicolumn{1}{r|}{36.64}                 &                                \\ \cline{4-10}
\multicolumn{1}{|l|}{}                                & \multicolumn{1}{c|}{}                            & \multicolumn{1}{c|}{}                               & \multicolumn{1}{c|}{9}  & 8,813,264,533,265                 & \multicolumn{1}{r|}{281.39}               & \multicolumn{1}{r|}{2.88}                 & \multicolumn{1}{r|}{139.046}                                                         & 0.109                          & \multicolumn{1}{r|}{39.14}                 &                                \\ \cline{4-10}
\multicolumn{1}{|l|}{}                                & \multicolumn{1}{c|}{}                            & \multicolumn{1}{c|}{}                               & \multicolumn{1}{c|}{10} & 89,563,892,212,629                & \multicolumn{1}{r|}{2,981.74}             & \multicolumn{1}{r|}{2.88}                 & \multicolumn{1}{r|}{2,262.99}                                                        & 0.109                          & \multicolumn{1}{r|}{41.64}                 &                                \\ \cline{4-10}
\multicolumn{1}{|l|}{}                                & \multicolumn{1}{c|}{}                            & \multicolumn{1}{c|}{}                               & \multicolumn{1}{c|}{11} & 822,551,101,011,469               & \multicolumn{1}{r|}{\textgreater 5 hours} & \multicolumn{1}{r|}{2.88}                 & \multicolumn{1}{r|}{\textgreater 5 hours}                                            & 0.109                          & \multicolumn{1}{r|}{44.14}                 &                                \\ \hline
\multicolumn{1}{|l|}{\multirow{8}{*}{com-orkut}}      & \multicolumn{1}{c|}{\multirow{8}{*}{3,072,441}}  & \multicolumn{1}{c|}{\multirow{8}{*}{117,185,083}}   & \multicolumn{1}{c|}{4}  & 3,221,946,137                     & \multicolumn{1}{r|}{3.10}                 & \multicolumn{1}{r|}{292.35}               & \multicolumn{1}{r|}{0.426}                                                           & 8.83                           & \multicolumn{1}{r|}{1,394.40}              & \multirow{8}{*}{1,400.83}      \\ \cline{4-10}
\multicolumn{1}{|l|}{}                                & \multicolumn{1}{c|}{}                            & \multicolumn{1}{c|}{}                               & \multicolumn{1}{c|}{5}  & 15,766,607,860                    & \multicolumn{1}{r|}{4.94}                 & \multicolumn{1}{r|}{385.04}               & \multicolumn{1}{r|}{1.014}                                                           & 13.869                         & \multicolumn{1}{r|}{1,399.40}              &                                \\ \cline{4-10}
\multicolumn{1}{|l|}{}                                & \multicolumn{1}{c|}{}                            & \multicolumn{1}{c|}{}                               & \multicolumn{1}{c|}{6}  & 75,249,427,585                    & \multicolumn{1}{r|}{12.57}                & \multicolumn{1}{r|}{462.05}               & \multicolumn{1}{r|}{3.506}                                                           & 17.229                         & \multicolumn{1}{r|}{1,404.40}              &                                \\ \cline{4-10}
\multicolumn{1}{|l|}{}                                & \multicolumn{1}{c|}{}                            & \multicolumn{1}{c|}{}                               & \multicolumn{1}{c|}{7}  & 353,962,921,685                   & \multicolumn{1}{r|}{42.09}                & \multicolumn{1}{r|}{517.29}               & \multicolumn{1}{r|}{11.719}                                                          & 20.331                         & \multicolumn{1}{r|}{1,394.40}              &                                \\ \cline{4-10}
\multicolumn{1}{|l|}{}                                & \multicolumn{1}{c|}{}                            & \multicolumn{1}{c|}{}                               & \multicolumn{1}{c|}{8}  & 1,632,691,821,296                 & \multicolumn{1}{r|}{150.87}               & \multicolumn{1}{r|}{559.75}               & \multicolumn{1}{r|}{45.319}                                                          & 26.137                         & \multicolumn{1}{r|}{1,414.40}              &                                \\ \cline{4-10}
\multicolumn{1}{|l|}{}                                & \multicolumn{1}{c|}{}                            & \multicolumn{1}{c|}{}                               & \multicolumn{1}{c|}{9}  & 7,248,102,160,867                 & \multicolumn{1}{r|}{584.39}               & \multicolumn{1}{r|}{598.88}               & \multicolumn{1}{r|}{212.912}                                                         & 33.644                         & \multicolumn{1}{r|}{1,419.40}              &                                \\ \cline{4-10}
\multicolumn{1}{|l|}{}                                & \multicolumn{1}{c|}{}                            & \multicolumn{1}{c|}{}                               & \multicolumn{1}{c|}{10} & 30,288,138,110,629                & \multicolumn{1}{r|}{2,315.89}             & \multicolumn{1}{r|}{647.18}               & \multicolumn{1}{r|}{1,002.165}                                                       & 39.957                         & \multicolumn{1}{r|}{1,424.40}              &                                \\ \cline{4-10}
\multicolumn{1}{|l|}{}                                & \multicolumn{1}{c|}{}                            & \multicolumn{1}{c|}{}                               & \multicolumn{1}{c|}{11} & 117,138,620,358,191               & \multicolumn{1}{r|}{8,843.51}             & \multicolumn{1}{r|}{647.18}               & \multicolumn{1}{r|}{4,421.597}                                                       & 48.101                         & \multicolumn{1}{r|}{1,429.40}              &                                \\ \hline
\multicolumn{1}{|l|}{\multirow{8}{*}{com-friendster}} & \multicolumn{1}{c|}{\multirow{8}{*}{65,608,366}} & \multicolumn{1}{c|}{\multirow{8}{*}{1,806,067,135}} & \multicolumn{1}{c|}{4}  & 8,963,503,263                     & \multicolumn{1}{r|}{109.46}               & \multicolumn{1}{r|}{Out of memory}        & \multicolumn{1}{r|}{10.215}                                                          & 44.897                         & \multicolumn{1}{r|}{21,209.19}             & \multirow{8}{*}{21,220.95}     \\ \cline{4-10}
\multicolumn{1}{|l|}{}                                & \multicolumn{1}{c|}{}                            & \multicolumn{1}{c|}{}                               & \multicolumn{1}{c|}{5}  & 21,710,817,218                    & \multicolumn{1}{r|}{111.75}               & \multicolumn{1}{r|}{Out of memory}        & \multicolumn{1}{r|}{11.796}                                                          & 53.874                         & \multicolumn{1}{r|}{21,215.44}             &                                \\ \cline{4-10}
\multicolumn{1}{|l|}{}                                & \multicolumn{1}{c|}{}                            & \multicolumn{1}{c|}{}                               & \multicolumn{1}{c|}{6}  & 59,926,510,355                    & \multicolumn{1}{r|}{115.52}               & \multicolumn{1}{r|}{Out of memory}        & \multicolumn{1}{r|}{17.22}                                                           & 63.874                         & \multicolumn{1}{r|}{21,221.69}             &                                \\ \cline{4-10}
\multicolumn{1}{|l|}{}                                & \multicolumn{1}{c|}{}                            & \multicolumn{1}{c|}{}                               & \multicolumn{1}{c|}{7}  & 296,858,496,789                   & \multicolumn{1}{r|}{139.98}               & \multicolumn{1}{r|}{Out of memory}        & \multicolumn{1}{r|}{45.697}                                                          & 66.544                         & \multicolumn{1}{r|}{21,209.19}             &                                \\ \cline{4-10}
\multicolumn{1}{|l|}{}                                & \multicolumn{1}{c|}{}                            & \multicolumn{1}{c|}{}                               & \multicolumn{1}{c|}{8}  & 3,120,447,373,827                 & \multicolumn{1}{r|}{300.62}               & \multicolumn{1}{r|}{Out of memory}        & \multicolumn{1}{r|}{99.866}                                                          & 67.064                         & \multicolumn{1}{r|}{21,234.19}             &                                \\ \cline{4-10}
\multicolumn{1}{|l|}{}                                & \multicolumn{1}{c|}{}                            & \multicolumn{1}{c|}{}                               & \multicolumn{1}{c|}{9}  & 40,033,489,612,826                & \multicolumn{1}{r|}{1,796.12}             & \multicolumn{1}{r|}{Out of memory}        & \multicolumn{1}{r|}{803.53}                                                          & 71.404                         & \multicolumn{1}{r|}{21,240.44}             &                                \\ \cline{4-10}
\multicolumn{1}{|l|}{}                                & \multicolumn{1}{c|}{}                            & \multicolumn{1}{c|}{}                               & \multicolumn{1}{c|}{10} & 487,090,833,092,739               & \multicolumn{1}{r|}{16,836.41}            & \multicolumn{1}{r|}{Out of memory}        & \multicolumn{1}{r|}{12,775.67}                                                       & 71.051                         & \multicolumn{1}{r|}{21,246.69}             &                                \\ \cline{4-10}
\multicolumn{1}{|l|}{}                                & \multicolumn{1}{c|}{}                            & \multicolumn{1}{c|}{}                               & \multicolumn{1}{c|}{11} & 5,403,375,502,221,430             & \multicolumn{1}{r|}{\textgreater 5hours}  & \multicolumn{1}{r|}{Out of memory}        & \multicolumn{1}{r|}{\textgreater 5hours}                                             & 71.448                         & \multicolumn{1}{r|}{21,252.94}             &                                \\ \hline
\multicolumn{1}{|l|}{\multirow{5}{*}{com-lj}}         & \multicolumn{1}{c|}{\multirow{5}{*}{3,997,962}}  & \multicolumn{1}{c|}{\multirow{5}{*}{34,681,189}}    & \multicolumn{1}{c|}{4}  & 5,216,918,441                     & \multicolumn{1}{r|}{1.77}                 & \multicolumn{1}{r|}{268.06}               & \multicolumn{1}{r|}{0.104}                                                           & 10.864                         & \multicolumn{1}{r|}{513.39}                & \multirow{5}{*}{499.47}        \\ \cline{4-10}
\multicolumn{1}{|l|}{}                                & \multicolumn{1}{c|}{}                            & \multicolumn{1}{c|}{}                               & \multicolumn{1}{c|}{5}  & 246,378,629,120                   & \multicolumn{1}{r|}{7.52}                 & \multicolumn{1}{r|}{1,475.99}             & \multicolumn{1}{r|}{0.943}                                                           & 68.966                         & \multicolumn{1}{r|}{524.02}                &                                \\ \cline{4-10}
\multicolumn{1}{|l|}{}                                & \multicolumn{1}{c|}{}                            & \multicolumn{1}{c|}{}                               & \multicolumn{1}{c|}{6}  & 10,990,740,312,954                & \multicolumn{1}{r|}{258.46}               & \multicolumn{1}{r|}{7,816.13}             & \multicolumn{1}{r|}{23.792}                                                          & 379.88                         & \multicolumn{1}{r|}{534.64}                &                                \\ \cline{4-10}
\multicolumn{1}{|l|}{}                                & \multicolumn{1}{c|}{}                            & \multicolumn{1}{c|}{}                               & \multicolumn{1}{c|}{7}  & 449,022,426,169,164               & \multicolumn{1}{r|}{10,733.21}            & \multicolumn{1}{r|}{\textgreater 5 hours} & \multicolumn{1}{r|}{1,077.66}                                                        & 1,639.537                      & \multicolumn{1}{r|}{513.39}                &                                \\ \cline{4-10}
\multicolumn{1}{|l|}{}                                & \multicolumn{1}{c|}{}                            & \multicolumn{1}{c|}{}                               & \multicolumn{1}{c|}{8}  & 16,890,998,195,437,600            & \multicolumn{1}{r|}{\textgreater 5hours}  & \multicolumn{1}{r|}{\textgreater 5 hours} & \multicolumn{1}{r|}{\textgreater 5 hours}                                            & 6,850.989                      & \multicolumn{1}{r|}{555.89}                &                                \\ \hline
\end{tabular}%

%% file: fig/3-implementation/oritentation.tex
\begin{tabular}{l|c|c|c|c|c|}
\cline{2-6}
                                      & Undirected & \multicolumn{2}{c|}{Degree Orientation} & \multicolumn{2}{c|}{Degeneracy Orientation} \\ \hline
\multicolumn{1}{|l|}{Graph} &
  $d_{max}$ &
  $d_{max}$ &
  \begin{tabular}[c]{@{}c@{}}Preprocessing \\ Time in seconds\end{tabular} &
  $d_{max}$ &
  \begin{tabular}[c]{@{}c@{}}Preprocessing \\ Time in seconds\end{tabular} \\ \hline
\multicolumn{1}{|l|}{as-skitter}     & 35,455     & 231               & 0.005               & 111                 & 0.205                 \\ \hline
\multicolumn{1}{|l|}{com-dblp}        & 343        & 113               & 0.002               & 113                 & 0.051                 \\ \hline
\multicolumn{1}{|l|}{com-orkut}       & 33,313     & 535               & 0.056               & 253                 & 0.833                 \\ \hline
\multicolumn{1}{|l|}{com-friendster}  & 5,214      & 868               & 5.465              & 304                 & 12.294                \\ \hline
\multicolumn{1}{|l|}{com-lj} & 14,815     & 524               & 0.016               & 360                 & 0.421                 \\ \hline
\end{tabular}%

%% file: fig/4-evaluation/tab-extended.tex
\begin{table*}[]
\caption{
A comparison of the total execution time (in seconds) between ARB-COUNT~\cite{arb-count} and Pivoter~\cite{powerpivoting} on two different CPUs, and our GPU implementations on three different GPUs
}
\label{tab:compare_all}
\resizebox{\textwidth}{!}{%

\begin{tabular}{lc|rr|rr|rr|rr|rr|}
\cline{3-12}
                                                      & \multicolumn{1}{l|}{} & \multicolumn{2}{l|}{Intel Xeon Scalable (60 Threads)}                    & \multicolumn{2}{l|}{Intel Xeon Platinum 8260 (96 Threads)}              & \multicolumn{2}{c|}{Nvidia Volta V100}                                                                                & \multicolumn{2}{c|}{Nvidia Ampere RTX3090}                                                           & \multicolumn{2}{c|}{Nvidia Ampere A100}                                                          \\ \hline
\multicolumn{1}{|c|}{Graph}                           & k                     & \multicolumn{1}{c|}{ARB-COUNT}            & \multicolumn{1}{c|}{Pivoter} & \multicolumn{1}{c|}{ARB-COUNT}           & \multicolumn{1}{c|}{Pivoter} & \multicolumn{1}{c|}{\begin{tabular}[c]{@{}c@{}}GPU-Graph\\ Orientation\end{tabular}} & \multicolumn{1}{c|}{GPU-Pivot} & \multicolumn{1}{r|}{\begin{tabular}[c]{@{}r@{}}GPU-Graph\\ Orientation\end{tabular}} & GPU-Pivot     & \multicolumn{1}{c|}{\begin{tabular}[c]{@{}c@{}}GPU-Graph\\ Orientation\end{tabular}} & GPU-Pivot \\ \hline
\multicolumn{1}{|l|}{\multirow{8}{*}{as-skitter}}     & 4                     & \multicolumn{1}{r|}{0.60}                 & 20.90                        & \multicolumn{1}{r|}{0.073}               & 32.667                       & \multicolumn{1}{r|}{0.034}                                                           & 0.459                          & \multicolumn{1}{r|}{0.026}                                                           & 0.525         & \multicolumn{1}{r|}{0.028}                                                           & 0.546     \\ \cline{2-12} 
\multicolumn{1}{|l|}{}                                & 5                     & \multicolumn{1}{r|}{0.67}                 & 22.54                        & \multicolumn{1}{r|}{0.127}               & 35.333                       & \multicolumn{1}{r|}{0.069}                                                           & 0.721                          & \multicolumn{1}{r|}{0.039}                                                           & 0.654         & \multicolumn{1}{r|}{0.034}                                                           & 0.682     \\ \cline{2-12} 
\multicolumn{1}{|l|}{}                                & 6                     & \multicolumn{1}{r|}{1.24}                 & 25.30                        & \multicolumn{1}{r|}{0.635}               & 37.000                       & \multicolumn{1}{r|}{0.245}                                                           & 1.01                           & \multicolumn{1}{r|}{0.172}                                                           & 0.896         & \multicolumn{1}{r|}{0.150}                                                           & 0.962     \\ \cline{2-12} 
\multicolumn{1}{|l|}{}                                & 7                     & \multicolumn{1}{r|}{5.73}                 & 27.14                        & \multicolumn{1}{r|}{5.28}                & 39.000                       & \multicolumn{1}{r|}{1.434}                                                           & 1.269                          & \multicolumn{1}{r|}{0.886}                                                           & 1.135         & \multicolumn{1}{r|}{0.729}                                                           & 1.314     \\ \cline{2-12} 
\multicolumn{1}{|l|}{}                                & 8                     & \multicolumn{1}{r|}{28.38}                & 27.46                        & \multicolumn{1}{r|}{39.699}              & 39.667                       & \multicolumn{1}{r|}{9.531}                                                           & 1.585                          & \multicolumn{1}{r|}{7.578}                                                           & 1.398         & \multicolumn{1}{r|}{5.804}                                                           & 1.731     \\ \cline{2-12} 
\multicolumn{1}{|l|}{}                                & 9                     & \multicolumn{1}{r|}{158.45}               & 28.45                        & \multicolumn{1}{r|}{-}                   & -                            & \multicolumn{1}{r|}{68.221}                                                          & 1.835                          & \multicolumn{1}{r|}{59.026}                                                          & 1.634         & \multicolumn{1}{r|}{46.297}                                                          & 1.967     \\ \cline{2-12} 
\multicolumn{1}{|l|}{}                                & 10                    & \multicolumn{1}{r|}{854.87}               & 28.45                        & \multicolumn{1}{r|}{-}                   & -                            & \multicolumn{1}{r|}{474.785}                                                         & 1.777                          & \multicolumn{1}{r|}{414.812}                                                         & 1.694         & \multicolumn{1}{r|}{351.133}                                                         & 2.217     \\ \cline{2-12} 
\multicolumn{1}{|l|}{}                                & 11                    & \multicolumn{1}{r|}{4,158.53}             & 28.45                        & \multicolumn{1}{r|}{-}                   & -                            & \multicolumn{1}{r|}{2,997.385}                                                       & 1.777                          & \multicolumn{1}{r|}{2,572.444}                                                       & 1.764         & \multicolumn{1}{r|}{2,493.202}                                                       & 2.15      \\ \hline
\multicolumn{1}{|l|}{\multirow{8}{*}{com-dblp}}       & 4                     & \multicolumn{1}{r|}{0.10}                 & 2.88                         & \multicolumn{1}{r|}{-}                   & -                            & \multicolumn{1}{r|}{0.008}                                                           & 0.109                          & \multicolumn{1}{r|}{0.009}                                                           & 0.118         & \multicolumn{1}{r|}{0.010}                                                           & 0.112     \\ \cline{2-12} 
\multicolumn{1}{|l|}{}                                & 5                     & \multicolumn{1}{r|}{0.13}                 & 2.88                         & \multicolumn{1}{r|}{-}                   & -                            & \multicolumn{1}{r|}{0.016}                                                           & 0.109                          & \multicolumn{1}{r|}{0.016}                                                           & 0.118         & \multicolumn{1}{r|}{0.015}                                                           & 0.112     \\ \cline{2-12} 
\multicolumn{1}{|l|}{}                                & 6                     & \multicolumn{1}{r|}{0.30}                 & 2.88                         & \multicolumn{1}{r|}{-}                   & -                            & \multicolumn{1}{r|}{0.042}                                                           & 0.109                          & \multicolumn{1}{r|}{0.031}                                                           & 0.118         & \multicolumn{1}{r|}{0.030}                                                           & 0.112     \\ \cline{2-12} 
\multicolumn{1}{|l|}{}                                & 7                     & \multicolumn{1}{r|}{2.05}                 & 2.88                         & \multicolumn{1}{r|}{-}                   & -                            & \multicolumn{1}{r|}{0.545}                                                           & 0.109                          & \multicolumn{1}{r|}{0.408}                                                           & 0.118         & \multicolumn{1}{r|}{0.444}                                                           & 0.112     \\ \cline{2-12} 
\multicolumn{1}{|l|}{}                                & 8                     & \multicolumn{1}{r|}{24.06}                & 2.88                         & \multicolumn{1}{r|}{-}                   & -                            & \multicolumn{1}{r|}{9.031}                                                           & 0.109                          & \multicolumn{1}{r|}{7.912}                                                           & 0.118         & \multicolumn{1}{r|}{8.835}                                                           & 0.112     \\ \cline{2-12} 
\multicolumn{1}{|l|}{}                                & 9                     & \multicolumn{1}{r|}{281.39}               & 2.88                         & \multicolumn{1}{r|}{-}                   & -                            & \multicolumn{1}{r|}{139.046}                                                         & 0.109                          & \multicolumn{1}{r|}{150.483}                                                         & 0.118         & \multicolumn{1}{r|}{159.794}                                                         & 0.112     \\ \cline{2-12} 
\multicolumn{1}{|l|}{}                                & 10                    & \multicolumn{1}{r|}{2,981.74}             & 2.88                         & \multicolumn{1}{r|}{-}                   & -                            & \multicolumn{1}{r|}{2,262.99}                                                        & 0.109                          & \multicolumn{1}{r|}{2,190.995}                                                       & 0.118         & \multicolumn{1}{r|}{2,435.950}                                                       & 0.112     \\ \cline{2-12} 
\multicolumn{1}{|l|}{}                                & 11                    & \multicolumn{1}{r|}{\textgreater 5 hours} & 2.88                         & \multicolumn{1}{r|}{-}                   & -                            & \multicolumn{1}{r|}{\textgreater 5 hours}                                            & 0.109                          & \multicolumn{1}{r|}{\textgreater 5 hours}                                            & 0.118         & \multicolumn{1}{r|}{\textgreater 5 hours}                                            & 0.112     \\ \hline
\multicolumn{1}{|l|}{\multirow{8}{*}{com-orkut}}      & 4                     & \multicolumn{1}{r|}{3.10}                 & 292.35                       & \multicolumn{1}{r|}{1.614}               & 308.667                      & \multicolumn{1}{r|}{0.426}                                                           & 8.83                           & \multicolumn{1}{r|}{0.328}                                                           & 8.534         & \multicolumn{1}{r|}{0.323}                                                           & 6.612     \\ \cline{2-12} 
\multicolumn{1}{|l|}{}                                & 5                     & \multicolumn{1}{r|}{4.94}                 & 385.04                       & \multicolumn{1}{r|}{2.863}               & 400.667                      & \multicolumn{1}{r|}{1.014}                                                           & 13.869                         & \multicolumn{1}{r|}{0.733}                                                           & 14.403        & \multicolumn{1}{r|}{0.606}                                                           & 10.926    \\ \cline{2-12} 
\multicolumn{1}{|l|}{}                                & 6                     & \multicolumn{1}{r|}{12.57}                & 462.05                       & \multicolumn{1}{r|}{8.694}               & 481.333                      & \multicolumn{1}{r|}{3.506}                                                           & 17.229                         & \multicolumn{1}{r|}{2.344}                                                           & 19.244        & \multicolumn{1}{r|}{1.827}                                                           & 14.046    \\ \cline{2-12} 
\multicolumn{1}{|l|}{}                                & 7                     & \multicolumn{1}{r|}{42.09}                & 517.29                       & \multicolumn{1}{r|}{33.278}              & 525.333                      & \multicolumn{1}{r|}{11.719}                                                          & 20.331                         & \multicolumn{1}{r|}{8.522}                                                           & 22.918        & \multicolumn{1}{r|}{7.681}                                                           & 17.321    \\ \cline{2-12} 
\multicolumn{1}{|l|}{}                                & 8                     & \multicolumn{1}{r|}{150.87}               & 559.75                       & \multicolumn{1}{r|}{133.695}             & 583.333                      & \multicolumn{1}{r|}{45.319}                                                          & 26.137                         & \multicolumn{1}{r|}{39.968}                                                          & 28.159        & \multicolumn{1}{r|}{30.002}                                                          & 22.483    \\ \cline{2-12} 
\multicolumn{1}{|l|}{}                                & 9                     & \multicolumn{1}{r|}{584.39}               & 598.88                       & \multicolumn{1}{r|}{-}                   & -                            & \multicolumn{1}{r|}{212.912}                                                         & 33.644                         & \multicolumn{1}{r|}{193.653}                                                         & 34.694        & \multicolumn{1}{r|}{140.732}                                                         & 29.82     \\ \cline{2-12} 
\multicolumn{1}{|l|}{}                                & 10                    & \multicolumn{1}{r|}{2,315.89}             & 647.18                       & \multicolumn{1}{r|}{-}                   & -                            & \multicolumn{1}{r|}{1,002.165}                                                       & 39.957                         & \multicolumn{1}{r|}{925.879}                                                         & 41.393        & \multicolumn{1}{r|}{668.427}                                                         & 38.218    \\ \cline{2-12} 
\multicolumn{1}{|l|}{}                                & 11                    & \multicolumn{1}{r|}{8,843.51}             & 647.18                       & \multicolumn{1}{r|}{-}                   & -                            & \multicolumn{1}{r|}{4,421.597}                                                       & 48.101                         & \multicolumn{1}{r|}{4,086.51}                                                        & 46.564        & \multicolumn{1}{r|}{3,150.510}                                                       & 44.815    \\ \hline
\multicolumn{1}{|l|}{\multirow{8}{*}{com-friendster}} & 4                     & \multicolumn{1}{r|}{109.46}               & Out of memory                        & \multicolumn{1}{r|}{70.01}               & 4,433.5                      & \multicolumn{1}{r|}{10.215}                                                          & 44.897                         & \multicolumn{1}{r|}{Out of memory}                                                   & Out of memory & \multicolumn{1}{r|}{9.126}                                                           & 33.202    \\ \cline{2-12} 
\multicolumn{1}{|l|}{}                                & 5                     & \multicolumn{1}{r|}{111.75}               & Out of memory                        & \multicolumn{1}{r|}{70.817}              & 4,489.5                      & \multicolumn{1}{r|}{11.796}                                                          & 53.874                         & \multicolumn{1}{r|}{Out of memory}                                                   & Out of memory & \multicolumn{1}{r|}{10.356}                                                          & 38.212    \\ \cline{2-12} 
\multicolumn{1}{|l|}{}                                & 6                     & \multicolumn{1}{r|}{115.52}               & Out of memory                        & \multicolumn{1}{r|}{74.418}              & 4,554.5                      & \multicolumn{1}{r|}{17.22}                                                           & 63.874                         & \multicolumn{1}{r|}{Out of memory}                                                   & Out of memory & \multicolumn{1}{r|}{14.391}                                                          & 47.318    \\ \cline{2-12} 
\multicolumn{1}{|l|}{}                                & 7                     & \multicolumn{1}{r|}{139.98}               & Out of memory                        & \multicolumn{1}{r|}{93.549}              & 4,537.5                      & \multicolumn{1}{r|}{45.697}                                                          & 66.544                         & \multicolumn{1}{r|}{Out of memory}                                                   & Out of memory & \multicolumn{1}{r|}{31.034}                                                          & 47.408    \\ \cline{2-12} 
\multicolumn{1}{|l|}{}                                & 8                     & \multicolumn{1}{r|}{300.62}               & Out of memory                        & \multicolumn{1}{r|}{385.024}             & 4,556.5                      & \multicolumn{1}{r|}{99.866}                                                          & 67.064                         & \multicolumn{1}{r|}{Out of memory}                                                   & Out of memory & \multicolumn{1}{r|}{74.857}                                                          & 47.07     \\ \cline{2-12} 
\multicolumn{1}{|l|}{}                                & 9                     & \multicolumn{1}{r|}{1,796.12}             & Out of memory                        & \multicolumn{1}{r|}{-}                   & -                            & \multicolumn{1}{r|}{803.53}                                                          & 71.404                         & \multicolumn{1}{r|}{Out of memory}                                                   & Out of memory & \multicolumn{1}{r|}{660.689}                                                         & 46.115    \\ \cline{2-12} 
\multicolumn{1}{|l|}{}                                & 10                    & \multicolumn{1}{r|}{16,836.41}            & Out of memory                        & \multicolumn{1}{r|}{-}                   & -                            & \multicolumn{1}{r|}{12,775.67}                                                       & 71.051                         & \multicolumn{1}{r|}{Out of memory}                                                   & Out of memory & \multicolumn{1}{r|}{11,911.473}                                                      & 45.224    \\ \cline{2-12} 
\multicolumn{1}{|l|}{}                                & 11                    & \multicolumn{1}{r|}{\textgreater 5hours}  & Out of memory                        & \multicolumn{1}{r|}{-}                   & -                            & \multicolumn{1}{r|}{\textgreater 5hours}                                             & 71.448                         & \multicolumn{1}{r|}{Out of memory}                                                   & Out of memory & \multicolumn{1}{r|}{\textgreater 5 hours}                                            & 44.311    \\ \hline
\multicolumn{1}{|l|}{\multirow{5}{*}{com-lj}}         & 4                     & \multicolumn{1}{r|}{1.77}                 & 268.06                       & \multicolumn{1}{r|}{0.416}               & 299.500                      & \multicolumn{1}{r|}{0.104}                                                           & 10.864                         & \multicolumn{1}{r|}{0.09}                                                            & 9.339         & \multicolumn{1}{r|}{0.093}                                                           & 8.529     \\ \cline{2-12} 
\multicolumn{1}{|l|}{}                                & 5                     & \multicolumn{1}{r|}{7.52}                 & 1,475.99                     & \multicolumn{1}{r|}{5.587}               & 1,500.000                    & \multicolumn{1}{r|}{0.943}                                                           & 68.966                         & \multicolumn{1}{r|}{0.725}                                                           & 72.087        & \multicolumn{1}{r|}{0.695}                                                           & 53.79     \\ \cline{2-12} 
\multicolumn{1}{|l|}{}                                & 6                     & \multicolumn{1}{r|}{258.46}               & 7,816.13                     & \multicolumn{1}{r|}{256.241}             & \textgreater 1 hour          & \multicolumn{1}{r|}{23.792}                                                          & 379.88                         & \multicolumn{1}{r|}{23.15}                                                           & 404.725       & \multicolumn{1}{r|}{18.697}                                                          & 301.772   \\ \cline{2-12} 
\multicolumn{1}{|l|}{}                                & 7                     & \multicolumn{1}{r|}{10,733.21}            & \textgreater 5 hours         & \multicolumn{1}{r|}{\textgreater 1 hour} & \textgreater 1 hour          & \multicolumn{1}{r|}{1,077.66}                                                        & 1,639.537                      & \multicolumn{1}{r|}{1,400.245}                                                       & 1,836.612     & \multicolumn{1}{r|}{952.845}                                                         & 1,396.37  \\ \cline{2-12} 
\multicolumn{1}{|l|}{}                                & 8                     & \multicolumn{1}{r|}{\textgreater 5hours}  & \textgreater 5 hours         & \multicolumn{1}{r|}{\textgreater 1 hour} & \textgreater 1 hour          & \multicolumn{1}{r|}{\textgreater 5 hours}                                            & 6,850.989                      & \multicolumn{1}{r|}{\textgreater 5 hours}                                            & 7,065.479     & \multicolumn{1}{r|}{\textgreater 5 hours}                                            & 5,467.176 \\ \hline
\end{tabular}%

}
\end{table*}

%% file: sec/5-related.tex
\section{Related Work}

\textbf{Graph Orientation Approach to Clique Counting.}
Graph orientation is a fundamental approach to avoiding redundant clique discovery~\cite{chiba1985arboricity, degreeOrdering1, kclist, arb-count, orderingheuristic, gianinazzi2021parallel, lonkar2021accelerating}.
To our knowledge, ARB-COUNT~\cite{arb-count} is the state-of-the-art parallel implementation of $k$-clique counting on CPUs based on graph orientation.
We implement the graph orientation approach for $k$-clique counting on the GPU and compare our performance with ARB-COUNT~\cite{arb-count}.

\textbf{Pivoting Approach to Clique Counting.}
Pivoter~\cite{powerpivoting} is a recent work on $k$-clique counting that is inspired by the classical pivoting idea of Bron-Kerbosh's maximal clique finding~\cite{bkpivoting}.
ARB-COUNT~\cite{arb-count} compares to Pivoter and shows that Pivoter is advantageous for large $k$ values.
Pivoter is implemented on the CPU.
We implement the pivoting approach for $k$-clique counting on the GPU and compare our performance with Pivoter~\cite{powerpivoting}.

\textbf{Maximal Clique Enumeration.}
Enumerating the maximal cliques in a graph has been extensively studied on CPUs~\cite{mcc1, mcc2, mcc3, mcc4} and GPUs~\cite{mcg1, lessons, mcg3, bitsets, mcg4, mcg5}.
The pivoting approach is inspired by techniques used in maximal clique enumeration.
Our work solves the $k$-clique counting problem, but our techniques can be extended to the maximal clique problem.
To the best of our knowledge, none of the GPU works on maximal clique use edge-centric parallelization, use binary encoding for the induced sub-graph, or use sub-warp partitioning.

\textbf{Triangle Counting.}
Many works perform triangle counting on the CPU~\cite{tccpu1, tccpu2, tccpu3, tccpu4} or the GPU~\cite{gputc100, tc1, tc2, tc3, tc4, tc5, tc6, tc7, tc8, tc9}.
A triangle is a 3-clique which is a special case of a $k$-clique.
Our implementation performs $k$-clique counting for any $k$ value.

\textbf{Generalized Sub-graph Matching.}
Many works perform generalized sub-graph matching on the CPU~\cite{escape, motif1, motif2, motif3, motif4} and the GPU~\cite{almasri2022parallel, pangolin, sub0, sub1, sub2, sub3, sub4, sub5}.
These frameworks search for an arbitrary $k$-vertex sub-graph and support different values of $k$.
Due to generalization, such sub-graph matching frameworks suffer from memory explosion or prolonged execution times.
Our implementation is specialized for $k$-cliques which are an important special case of a $k$-vertex sub-graph.
Specializing for $k$-cliques enables optimizations that are not applicable to general sub-graphs, thereby providing better scalability for large values of $k$.

\textbf{Truss Decomposition.}
Recently, $k$-truss decomposition has received significant attention on CPUs~\cite{cputruss0, cputruss1, cputruss2, cputruss3} and GPUs~\cite{truss0, truss1, truss2, truss3, truss4, truss5}.
A $k$-truss is a relaxation of a $k$-clique.
Our work solves the $k$-clique counting problem.

\change{
\textbf{List Intersections.}
List intersection is an important operation at the heart of many sub-graph search algorithms.
Different list intersection strategies have been proposed for GPUs such as pointer-chasing~\cite{truss0}, binary-search~\cite{tc1}, merge-path~\cite{tc8}, hash-based~\cite{tc3}, tile-based~\cite{truss1}, bitmap-based~\cite{tc6}, and others.
Our implementations use binary-search intersections for extracting induced sub-graphs, and binary encoding for the lists in the induced sub-graph.

\textbf{Parallel Search Tree Traversal on GPUs.}
Parallel search tree traversal on GPUs has been studied for other graph problems such as minimum vertex cover~\cite{yamout2022parallel}.
This work uses a global worklist~\cite{kerbl2018broker} for dynamic load balancing because the search tree is narrow and highly imbalanced.
Our work shows that for the $k$-clique counting problem, edge-centric parallelization is sufficient for achieving reasonable load balance.
}

%% file: sec/6-conculsion.tex
\section{Conclusion}

We present parallel GPU implementations of $k$-clique counting that support both the graph orientation and pivoting approaches for eliminating redundant clique discovery.
We explore vertex-centric and edge-centric parallelization schemes and apply various optimizations such as binary encoding and sub-warp partitioning to reduce memory consumption and efficiently utilize parallel execution resources.
To the best of our knowledge, our work is the first GPU solution specialized for $k$-clique counting.

Our evaluation shows that our best GPU implementation substantially outperforms the best state-of-the-art parallel CPU implementation.
Our efficient memory management strategies enable us to process very large graphs with billions of edges for arbitrary values of $k$.
We also analyze the trade-offs between parallelization schemes, and show that the binary encoding and sub-warp partitioning optimizations yield significant performance gains.

%% file: sec/ack.tex
\begin{acks}

The authors would also like to thank the anonymous reviewers for their constructive feedback during the review process.
This work was supported in part by the IBM-Illinois Center for Cognitive Computing Systems Research (C3SR).

\end{acks}